\documentclass[aip, pop, reprint,numerical]{revtex4-1}

\usepackage{amsmath}				%	Collection of document classes and packages developed by the AMS
\usepackage{amssymb}				%	Collection of mathematical symbols
\usepackage{mathrsfs}				%	Ralph Smith's Formal Script Font. Works with \mathscr
\usepackage[mathscr]{euscript}		%	For nice fonts with \mathscr
\usepackage{amsthm}					% 	For theorems and proofs
\usepackage{xspace}					% Decides if a space should be put after a certain command. Useful for inline eqs.
\usepackage{graphicx}
\usepackage[caption = false]{subfig}	
\usepackage[utf8]{inputenc}			% Latin symbols
\usepackage{enumitem}				% Provides user control over the layout of enumerate, itemize, and description.
\usepackage{color} 					%	Text with color
\usepackage{verbatim}				% for comments. Use as .... "\verb| hello. My name is daniel. |" It is a typewritter font.
\usepackage{tensor}					% For tensor indices
\usepackage{easybmat}				% For tables
\usepackage{slashed}				% For Feynmann slashed notation

\usepackage{multirow}				%	For tables
\usepackage{longtable}				%	Longtable lets you have tables that span multiple pages.
\usepackage{booktabs}				% For nicer tables than the standard LaTEX tables

\newcommand{\eq}[1]{(\ref{#1})}
\newcommand{\Eq}[1]{Eq.~(\ref{#1})}
\newcommand{\Eqs}[1]{Eqs.~(\ref{#1})}
\newcommand{\Fig}[1]{Fig.~\ref{#1}}

\newcommand{\Sec}[1]{Sec.~\ref{#1}}
\newcommand{\Refa}[1]{Ref.~\onlinecite{#1}}
\newcommand{\Refs}[1]{Refs.~\onlinecite{#1}}
\newcommand{\App}[1]{Appendix~\ref{#1}}

\newcommand{\eg}{{e.g.,\/}\xspace}
\newcommand{\ie}{{i.e.,\/}\xspace}

\newcommand{\pd}{\partial}
\newcommand{\del }{\vec{\nabla}}

\newcommand{\dm}{\mathrm{d}}
\newcommand{\ep}{\epsilon}

\newcommand{\M}{\mathsf{M}_{\rm shell} }
\newcommand{\T}{\Theta}
\newcommand{\UR}{U_R}
\newcommand{\UT}{U_\Theta}
\newcommand{\PR}{P_R}
\newcommand{\PT}{P_\Theta}

\newcommand{\sft}{\mathsf{t}}
\newcommand{\sfm}{\mathsf{m}}
\newcommand{\sfX}{\mathsf{X}}
\newcommand{\sfR}{\mathsf{R}}
\newcommand{\sfT}{\mathsf{\Theta}}
\newcommand{\sfUR}{\mathsf{U}_{\mathsf{R}}}
\newcommand{\sfUT}{\mathsf{U}_{\mathsf{\Theta}}}

\newcommand{\sfsig}{\mathsf{\rho_S}}
\newcommand{\sfS}{\mathsf{S}}

\newcommand{\sfV}{\mathsf{V}}
\newcommand{\sfp}{\mathsf{p}}

\newcommand{\barR}{\bar{R}}
\newcommand{\barT}{\bar{\Theta}}
\newcommand{\barUR}{\bar{U}_R}

\newcommand{\barsig}{\bar{\sigma}}

\newcommand{\barL}{\bar{L}}
\newcommand{\barH}{\bar{H}}

\newcommand{\tR}{\widetilde{R}}
\newcommand{\tT}{\widetilde{\Theta}}
\newcommand{\tUR}{\widetilde{U}_R}
\newcommand{\tUT}{\widetilde{U}_\Theta}
\newcommand{\tsig}{\widetilde{\sigma}}

\newcommand{\tL}{\widetilde{L}}
\newcommand{\tH}{\widetilde{H}}

\newcommand{\hatR}{\widehat{R}_\ell}
\newcommand{\hatT}{\widehat{\Theta}_\ell}
\newcommand{\hatUR}{\widehat{U}_{R,\ell}}
\newcommand{\hatUT}{\widehat{U}_{\Theta,\ell}}
\newcommand{\hatsig}{\widehat{\sigma}_\ell}
\newcommand{\hatPsi}{\widehat{\Psi}_\ell}
\newcommand{\hatalpha}{\widehat{\alpha}_\ell}
\newcommand{\hatbeta}{\widehat{\beta}_\ell}

\newcommand{\sumell}{\sum_{\ell=1}^\infty}
\newcommand{\const}{{\rm const.}}

\newcommand{\mc}[1]{\mathcal{#1}}

\renewcommand{\vec}[1]{{\boldsymbol{\rm #1}}}

\usepackage{hyperref}
\hypersetup{
  colorlinks   	= true, 	%Colours links instead of ugly boxes
  urlcolor     	= blue, 	%Colour for external hyperlinks
  linkcolor    	= blue, 	%Colour of internal links
  citecolor   	= blue 	%Colour of citations
}
\begin{document}

\title{Degradation of performance in ICF implosions due to Rayleigh—Taylor instabilities:  a Hamiltonian perspective}

\author{D.~E.~Ruiz}
\email{deruiz@sandia.gov}
\affiliation{Sandia National Laboratories, P.O. Box 5800, Albuquerque, New Mexico 87185-1186, USA}

\date{\today}

%%%%%%%%%%%%%%%%%%%%%%%%%%%%%%%%%%%%%%%%%%%%%%%%%
%%%%%%%%%%%%%%%%%%%%%%%%%%%%%%%%%%%%%%%%%%%%%%%%%
%%%%%%%%%%%%%%%%%%%%%%%%%%%%%%%%%%%%%%%%%%%%%%%%%

\begin{abstract}

%Interfacial Rayleigh--Taylor (RT) instabilities are ubiquitous in inertial-confinement-fusion (ICF) implosions and are an important limiting factor of ICF performance.  In this work, we obtain a first-principle variational theory that describes an imploding spherical shell undergoing RT instabilities.  The model is based on a thin-shell approximation and includes the dynamical coupling between the imploding spherical shell and an adiabatically compressed fluid within its interior.  Based on the derived Hamiltonian framework, degradation trends of key ICF performance metrics (e.g., stagnation pressure, residual kinetic energy, and aerial density) are identified in terms of the initial RT-instability parameters (initial amplitude and Legendre mode).  Nonlinear results are obtained by numerically integrating the governing equations of this reduced model.

The Rayleigh--Taylor instability (RTI) is an ubiquitous phenomenon that occurs in inertial-confinement-fusion (ICF) implosions and is recognized as an important limiting factor of ICF performance. To analytically understand the RTI dynamics and its impact on ICF capsule implosions, we develop a first-principle variational theory that describes an imploding spherical shell undergoing RTI.  The model is based on a thin-shell approximation and includes the dynamical coupling between the imploding spherical shell and an adiabatically compressed fluid within its interior.  Using a quasilinear analysis, we study the degradation trends of key ICF performance metrics (e.g., stagnation pressure, residual kinetic energy, and aerial density) as functions of initial RTI parameters (e.g., the initial amplitude and Legendre mode), as well as the 1D implosion characteristics (e.g., the convergence ratio).  We compare analytical results from the theory against nonlinear results obtained by numerically integrating the governing equations of this reduced model.  Our findings emphasize the need to incorporate polar flows in the calculation of residual kinetic energy and demonstrate that higher convergence ratios in ICF implosions lead to significantly greater degradation of key performance metrics.

\end{abstract}

\maketitle

%%%%%%%%%%%%%%%%%%%%%%%%%%%%%%%%%%%%%%%%%%%%%%%%%
%%%%%%%%%%%%%%%%%%%%%%%%%%%%%%%%%%%%%%%%%%%%%%%%%
%%%%%%%%%%%%%%%%%%%%%%%%%%%%%%%%%%%%%%%%%%%%%%%%%
\section{Introduction}
\label{sec:intro}

One of the main approaches to produce thermonuclear self-heating plasmas in the laboratory has been laser-driven inertial confinement fusion (ICF).\cite{nuckolls1972,Lindl:2004jz} In this approach, powerful lasers directly or indirectly energize the outside surface of spherical capsules to achieve high ablation pressures of $\mc{O}(100~\mathrm{Mbar})$ and implosion velocities greater than 400~km/s. The resulting implosion compresses deuterium--tritium (DT) fusion fuel within its interior and heats it to relevant thermonuclear conditions, where DT fusion reactions can occur.  Recent experiments at the National Ignition Facility (NIF) have demonstrated hot-spot ignition using indirect drive.\cite{Ignition2022,theindirectdriveicfcollaboration2024}

To maximize the performance of an ICF implosion, high levels of symmetry are required to obtain high pressures, high areal densities, and high efficient energy coupling between the compressed fuel and the imploding shell.\cite{hurricane2016a,casey2023}  In this regard, low-mode asymmetries have been recognized as a major degradation mechanism limiting the performance of ICF implosions.  Due to the importance of this subject, many works have investigated the relationship between implosion asymmetries and performance degradation.  From the experimental standpoint, considerable effort has been put into understanding the effects of residual flows,\cite{rinderknecht2020,schlossberg2021,mannion2021b,woo2022} initial asymmetries in the shell thickness,\cite{casey2021} and asymmetries caused by the external drive.\cite{rinderknecht2020,schlossberg2021,mannion2021b,casey2023}  Such experimental work has been complemented by extensive numerical studies based on radiation-hydrodynamic simulations.\cite{scott2013,spears2014,kritcher2014,clark2016,bose2017,bose2018,woo2018a,woo2018,colaitis2022,nora2023}  Such works have reported significant results including the relationship between residual flows caused by $\ell=1$ mode asymmetries and the observed changes in the measured ion temperatures,\cite{spears2014} as well as the connection between the concept of residual kinetic energy and the decline of performance in ICF implosions.\cite{kritcher2014,bose2017,woo2018a}  Recent theoretical studies utilizing classical-mechanics models have provided valuable insights to this topic, elucidating the interplay between residual kinetic energy, residual flows, asymmetries in the areal density, and their impact on performance,\cite{bose2017,woo2018a,hurricane2020,hurricane2022,casey2023} as well as their influence on the initiation of the ignition process.\cite{woo2021a}  Based on this previous body of work, there is no question that low-mode asymmetries degrade ICF performance.  

Building upon the aforementioned investigations into low-mode asymmetries, this study introduces a theoretical model aimed at examining the connection between the amplification of such asymmetries through the well-established paradigm of the acceleration-driven Rayleigh--Taylor instability (RTI)\cite{rayleigh1883,taylor1950} and the subsequent impact on ICF capsule implosions. In this study, we focus on the scenario of an imploding spherical shell that decelerates as it converges onto an adiabatically compressed fluid within its interior. The work presented in this paper does not consider the acceleration phase of an ICF capsule implosion, where the external x-ray drive and the associated temporal variations, or ``swings", in drive asymmetry determine the initial conditions of the shell prior to deceleration.\cite{kritcher2021,kritcher2022a}

To analyze the problem considered, we adopt the thin-shell approximation\cite{ott1972} and introduce a variational framework for its dynamics. Notably, our Hamiltonian framework inherently incorporates the relationship between residual kinetic energy and the deterioration of metrics associated with the stagnation event.  To understand the degradation of ICF implosions, we numerically solve the resulting set of nonlinear equations, investigating the dependence on RTI-related parameters such as the initial amplitude and Legendre mode, as well as the implosion characteristics including the convergence ratio. Additionally, to gain further insight into the underlying dynamics, we derive a \textit{quasilinear} model that allows us to construct explicit expressions for the residual flow velocity, mean areal density, and residual kinetic energy as functions of asymmetries in the shell areal density. Finally, we compare the results of our theoretical predictions to the numerical calculations obtained from solving the fully nonlinear equations.

This work is organized as follows.  In \Sec{sec:notation}, we introduce the basic notation used throughout this work.  In \Sec{sec:basic}, we derive the governing equations describing a decelerating spherical shell. In \Sec{sec:VP}, we present a variational Hamiltonian formulation of the model, obtain the corresponding Euler--Lagrange equations, and discuss the conservation properties of the model.  In \Sec{sec:back}, we present analytical results for the simplified case of a 1D spherical implosion with no RTI.  In \Sec{sec:nonlinear}, we numerically solve the governing equations and provide initial insights on how RTI-induced asymmetries deteriorate the performance of ICF implosions  In \Sec{sec:quasi}, we present a quasilinear model that includes the two-way feedback between the background implosion dynamics of the spherical shell and the RTI growth.  In \Sec{sec:discussion}, we present the results of nonlinear calculations, examining the interplay between RTI and the degradation of ICF implosions. Additionally, we compare our numerical findings to the theoretical estimates obtained from the quasilinear model.  In \Sec{sec:conclusions}, we present our conclusions and suggest future research directions.  In \App{app:var}, we briefly demonstrate the mathematical procedure for taking variations of the action.  In \App{app:numerics}, we present details of our numerical solver of the governing equations.  In \App{app:quasi}, we present intermediate results needed to obtain the quasilinear model.

%%%%%%%%%%%%%%%%%%%%%%%%%%%%%%%%%%%%%%%%%%%%%%%%%
%%%%%%%%%%%%%%%%%%%%%%%%%%%%%%%%%%%%%%%%%%%%%%%%%
%%%%%%%%%%%%%%%%%%%%%%%%%%%%%%%%%%%%%%%%%%%%%%%%%
\section{Notation}
\label{sec:notation}

The following notation is used throughout the paper. The symbol ``$\doteq$'' denotes definitions.  Mathematical symbols written \textit{sans serif}, e.g.~$\sfR$, denote quantities with dimensional units attached to them while symbols written \textit{avec serif}, e.g.~$R$, denote dimensionless (or normalized) quantities.  In this work, ``barred" symbols denote unperturbed variables that are independent of the polar angle, e.g.~$\bar{R}$. Symbols with a ``tilde" denote quantities correspoding to RTI perturbations, e.g.~$\smash{\tR}$.  Finally, symbols with a ``hat" denote quantities characterizing their amplitude, e.g.~$\smash{\widehat{R}}$. Finally, in the Euler-Lagrange equations (ELEs), the notation ``$\delta a: $'' denotes that the corresponding equation was obtained by extremizing the action integral $\Lambda$ with respect to $a$.

%%%%%%%%%%%%%%%%%%%%%%%%%%%%%%%%%%%%%%%%%%%%%%%%%
%%%%%%%%%%%%%%%%%%%%%%%%%%%%%%%%%%%%%%%%%%%%%%%%%
%%%%%%%%%%%%%%%%%%%%%%%%%%%%%%%%%%%%%%%%%%%%%%%%%
\section{Basic equations}
\label{sec:basic}

%%%%%%%%%%%%%%%%%%%%%%%%%%%%%%%%%%%%%%%%%%%%%%%%%
\subsection{Shell kinematics}
\label{sec:kinematics}

\begin{figure}
	\begin{center}
	\includegraphics[scale=.6]{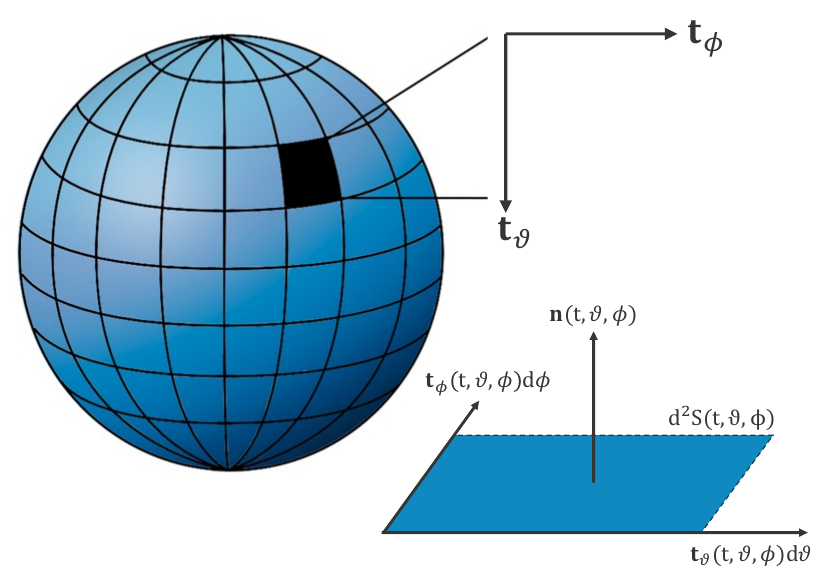}
	\end{center}
	\caption{Top: Parameterization of a spherical surface in the Lagrangian-label space $(\vartheta,\phi)$. Bottom right: Schematic diagram showing the relationship between the introduced two tangent vectors $(\vec{\mathsf{t}}_\vartheta,\vec{\mathsf{t}}_\phi)$, and the surface area differential $\dm^2 S$ and its normal $\vec{n}$.}
	\label{fig:schematic}
\end{figure}

In a similar manner to Ott's model,\cite{ott1972} we investigate the dynamics of an imploding shell of finite mass but with infinitesimal thickness.  This approximation is valid for perturbations whose characteristic wavelength $\lambda$ is much larger than the characteristic in-flight thickness of the shell.  To parameterize the surface of the shell, we introduce two independent Lagrangian coordinates $\vartheta$ and $\phi$, where  $\vartheta$ is the Lagrangian label corresponding to the polar angle $\theta$, and $\phi$ is the Lagrangian label corresponding to the azimuthal angle $\varphi$.  We consider the shell to be symmetric in the azimuthal direction.  The position vector $\vec{\sfX}=\vec{\sfX}(\sft,\vartheta,\phi)$ in the laboratory frame of a fluid parcel located within the interval $(\vartheta,\vartheta +\dm \vartheta)$ and $(\phi,\phi+\dm \phi)$ in the Lagrangian-label space is written as
\begin{equation}
	\vec{\sfX}(\sft,\vartheta,\phi)
		 \doteq \sfR(\sft,\vartheta) \,
				\vec{e}_r \boldsymbol{(} \sfT(\sft,\vartheta),\phi \boldsymbol{)},
	\label{eq:basic:X}
\end{equation}
where $\sfR=\sfR(\sft,\vartheta)$ and $\sfT=\sfT(\sft,\vartheta)$ are the time-dependent radial distance and polar angle of the fluid element in the laboratory frame. Due to the assumed azimuthal symmetry, the functions $\sfR$ and $\sfT$ do not depend on the Lagrangian label $\phi$.  In the spherical coordinate system, $(\vec{e}_r,\vec{e}_\theta,\vec{e}_\varphi)$ are the orthogonal unit vectors in the direction of increasing $r$, $\theta$, and $\varphi$, respectively.  In terms of the Cartesian unit vectors $(\vec{e}_x,\vec{e}_y,\vec{e}_z)$, they are written as follows:
\begin{align}
	\vec{e}_r(\theta,\varphi) 
			& \doteq \sin(\theta) \cos(\varphi) \vec{e}_x + \sin(\theta) \sin(\varphi) \vec{e}_y + \cos(\theta) \vec{e}_z, \\
	\vec{e}_\theta(\theta,\varphi) 
			& \doteq \frac{\pd \vec{e}_r}{\pd \theta} \notag \\
			& = \cos(\theta) \cos(\varphi) \vec{e}_x + \cos(\theta) \sin(\varphi) \vec{e}_y - \sin(\theta) \vec{e}_z, \\
	\vec{e}_\varphi(\theta,\varphi) 
			& \doteq \frac{1}{\sin(\theta)}\frac{\pd \vec{e}_r}{\pd \varphi} 
			 = -\sin(\varphi) \vec{e}_x + \cos(\varphi) \vec{e}_y.
\end{align}

With the parameterization in \Eq{eq:basic:X}, we define two tangent vectors along the shell surface (see \Fig{fig:schematic}):
\begin{align}
	\vec{\sft}_\vartheta(\sft,\vartheta,\phi) & \doteq \frac{\pd \vec{\sfX}}{\pd \vartheta}
				=	\frac{\pd \sfR}{\pd \vartheta} \, \vec{e}_r(\sfT,\phi) 
					+ \sfR \frac{\pd \sfT}{\pd \vartheta} \, \vec{e}_\theta(\sfT,\phi),  
		\label{eq:basic:t1} \\
	\vec{\sft}_\phi(\sft,\vartheta,\phi) & \doteq \frac{\pd \vec{\sfX}}{\pd \phi}
				=	\sfR \sin (\sfT) \, \vec{e}_\varphi(\phi).	
		\label{eq:basic:t2}		
\end{align}
The tangent vectors $\vec{\sft}_\vartheta$ and $\vec{\sft}_\phi$ are not necessarily unit vectors.  As illustrated by \Fig{fig:schematic}, the surface differential $\dm^2 \sfS(\sft,\vartheta,\phi)$ of a fluid parcel located within the interval $(\vartheta,\vartheta +\dm \vartheta)$ and $(\phi,\phi+\dm \phi)$ in the Lagrangian-label space is given by
\begin{equation}
	\dm^2 \sfS 	 \doteq \frac{\pd^2 \sfS}{\pd \vartheta \pd \phi} \, \dm \vartheta  \dm \phi,
	\label{eq:basic:d2S}	
\end{equation}
where
\begin{align}
	\frac{\pd^2 \sfS}{\pd \vartheta \pd \phi}
			\doteq	| \vec{\sft}_\vartheta \times \vec{\sft}_\phi | 
			=	\bigg[ \left( \sfR \frac{\pd \sfT}{\pd \vartheta} \right)^2 
							+ \left( \frac{\pd \sfR}{\pd \vartheta} \right)^2 
					\bigg]^{1/2} \sfR \sin(\sfT)
	\label{eq:basic:dS}
\end{align}
is the surface-area coefficient.  The unit vector $\vec{n} = \vec{n}(\sft,\vartheta,\phi)$ normal to the shell surface with Lagrangian parameters $\vartheta$ and $\phi$ is
\begin{align}
	\vec{n}(\sft,\vartheta,\phi) 	
		&\doteq \frac{\vec{\mathsf{t}}_\vartheta \times \vec{\mathsf{t}}_\phi}
					{ |\vec{\mathsf{t}}_\vartheta \times \vec{\mathsf{t}}_\phi |} \notag \\
		&=  \frac{\sfR \sin(\sfT)}{\pd^2 \sfS/\pd \vartheta \pd \phi} 
			\left[ \sfR \frac{\pd \sfT}{\pd \vartheta} \vec{e}_r(\sfT,\phi) - \frac{\pd \sfR}{\pd \vartheta} \vec{e}_\theta(\sfT,\phi) \right].
	\label{eq:basic:n}
\end{align}
Typically, $\vec{n}$ faces radially outwards:  for a perfect spherical shell, the radius only depends on time so that $\sfR=\sfR(\sft)$ and $\sfT=\vartheta$, which gives $\vec{n}=\vec{e}_r$.

With these preliminary results, we may now write the governing equation for the shell-implosion dynamics.  Let $\dm \sfm (\sft,\vartheta,\phi) \doteq \sfsig(\sft,\vartheta)\, \dm^2 \sfS$ be the infinitesimal mass of a fluid parcel located with Lagrangian label $(\vartheta,\phi)$.  Here $\sfsig = \sfsig(\sft,\vartheta)$ is a mass density per unit area, \ie the areal density, which is independent of the Lagrangian label $\phi$ due to the assumed azimuthal symmetry.  In this work, we consider the deceleration phase of the implosion.  Therefore, no external pressure sources that accelerate the shell inwards are considered.  However, the imploding shell is decelerated by a time-dependent pressure source $\sfp(\sft)$, which corresponds to the pressure of the fluid fuel located inside the cavity of the shell and increases as the shell is imploding.  In this case, the force differential $\dm \vec{\mathsf{F}}$ acting on a fluid parcel with Lagrangian label $(\vartheta,\phi)$ is 
\begin{equation}
	\dm \vec{\mathsf{F}}(\sft,\vartheta,\phi) 
			\doteq 	\sfp(\sft) \, \vec{n}(\sft,\vartheta,\phi) \, \dm^2 \sfS 
			= 		\sfp(\sft) (\vec{\sft}_\vartheta\times \vec{\sft}_\phi) \, 
					\dm \vartheta \, \dm \phi.
	\label{eq:basic:dF}
\end{equation}
We then invoke Newton's second law and obtain the following equation of motion for a fluid parcel with Lagrangian label $(\vartheta,\phi)$:
\begin{equation}
	\frac{\pd}{\pd \sft} \left( \dm \sfm \frac{\pd \vec{\sfX}}{\pd \sft} \right) 
			= \dm \vec{\mathsf{F}}.
	\label{eq:basic:Newton}
\end{equation}
When taking the time derivative of the position vector $\vec{\sfX}$, we find
\begin{equation}
	\frac{\pd }{\pd \sft}\vec{\sfX}(\sft,\vartheta,\phi) 
		= 		\frac{\pd \sfR}{\pd \sft} \vec{e}_r(\sfT,\phi) 
			+ 	\sfR \frac{\pd \sfT}{\pd \sft} \vec{e}_\theta(\sfT,\phi).
	\label{eq:basic:vel}
\end{equation}
We then introduce the shell velocity field $\vec{\mathsf{U}}(\sft,\vartheta,\phi) \doteq \sfUR(\sft,\vartheta) \, \vec{e}_r(\sfT,\phi) + \sfUT(\sft,\vartheta) \, \vec{e}_\theta(\sfT,\phi)$, where $\sfUR$ and $\sfUT$ are the associated radial and polar-angle velocity fields, respectively.  Hence, the governing equations for the variables $\sfR$, $\sfT$, $\sfUR$, and $\sfUT$ are
\begin{gather}
	\frac{\pd}{\pd \sft} \sfR = \sfUR, 
		\label{eq:basic:Raux}	\\
	\sfR\frac{\pd}{\pd \sft} \sfT = \sfUT, 
		\label{eq:basic:Thetaaux}\\
	\frac{\pd}{\pd \sft} \left( \sfsig \frac{\pd^2 \sfS}{\pd \vartheta \pd \phi} \, \sfUR \right) 
			= 		\sfp \sfR^2 \sin(\sfT) \frac{\pd \sfT}{\pd \vartheta}
				+ 	\sfsig \frac{\pd^2 \sfS}{\pd \vartheta \pd \phi} \frac{\sfUT^2}{\sfR}, 
		\label{eq:basic:Uaux}\\
	\frac{\pd}{\pd \sft} \left( \sfsig \frac{\pd^2 \sfS}{\pd \vartheta \pd \phi} \, \sfUT \right) 
			= - \sfp \sfR \sin(\sfT) \frac{\pd \sfR}{\pd \vartheta}
				- 	\sfsig \frac{\pd^2 \sfS}{\pd \vartheta \pd \phi} \frac{\sfUR \sfUT}{\sfR}.		
		\label{eq:basic:Vaux}
\end{gather}
These equations describe the kinematics of a decelerating spherical shell and are analagous to those found in  \Refs{taguchi1995,book1996}.  In contrast to other models describing asymmetric shell implosions,\cite{springer2018} these equations include radial and polar flows.  In this regard, the second term on the right-hand side of \Eq{eq:basic:Uaux} represents a centrifugal force that drives the shell material radially outward once the fluid parcel acquires polar angular momentum. This effect, which is not present in 1D models of spherical implosions, plays a significant role in inhibiting the radial compression of the shell.

%%%%%%%%%%%%%%%%%%%%%%%%%%%%%%%%%%%%%%%%%%%%%%%%%
\subsection{Shell areal density}
\label{sec:mass_density}

To obtain the governing equation for the shell areal density $\sfsig$, we note that the total mass of the imploding shell is given by
\begin{equation}
	\M \doteq \int  \sfsig\, \dm^2 \sfS 
		= 2 \pi \int_0^\pi \sfsig \,\frac{\pd^2 \sfS}{\pd \vartheta \pd \phi} \,
				 \dm \vartheta .
	\label{eq:basic:M}
\end{equation}
When assuming that the total mass of the shell is conserved, we have $\dm \M / \dm \sft =0$.  This leads to
\begin{equation}
	\frac{\pd }{\pd \sft} \left( \sfsig \frac{\pd^2 \sfS}{\pd \vartheta \pd \phi}  \right) = 0
	\label{eq:basic:m}
\end{equation}
so that
\begin{equation}
	\sfsig(\sft,\vartheta) \frac{\pd^2 \sfS}{\pd \vartheta \pd \phi} (\sft,\vartheta) 
		=	\mathsf{\rho}_{\mathsf{S},0}(\vartheta) \frac{\pd^2 \sfS_0}{\pd \vartheta \pd \phi}(\vartheta),
	\label{eq:basic:sigma}
\end{equation}
where $\mathsf{\rho}_{\mathsf{S},0}(\vartheta)\doteq\sfsig(0,\phi)$ is the initial mass density per unit area and
\begin{equation}
	\frac{\pd \sfS_0}{\pd \vartheta \pd \phi}(\vartheta)
		\doteq \frac{\pd \sfS}{\pd \vartheta \pd \phi}(0,\vartheta)
\end{equation}
is the surface-area coefficient \eq{eq:basic:dS} at time $\sft=0$.

Equation~\eq{eq:basic:sigma} is simple to understand: the mass differential $\dm \sfm = \sfsig \dm^2 \sfS$ of a fluid parcel with Lagrangian labels $(\vartheta,\phi)$ must be conserved throughout the implosion.  Thus, when the surface-area differential $\dm^2 \sfS$ locally expands or contracts during the implosion, the corresponding mass density $\sfsig$ must adjust to conserve the mass of the fluid parcel.

%%%%%%%%%%%%%%%%%%%%%%%%%%%%%%%%%%%%%%%%%%%%%%%%%
\subsection{Compressed-fuel pressure}
\label{sec:pressure}

In this work, we assume that the fuel is adiabatically compressed so that the fuel pressure obeys
\begin{equation}
	\sfp(\sft) = \sfp_0 \frac{\sfV^\gamma(0)}{\sfV^\gamma(\sft)} ,
	\label{eq:basic:p}
\end{equation}
where $\sfp_0$ is the fuel pressure when deceleration begins and $\gamma$ is the polytropic index of the fuel.  The time-dependent function $\sfV = \sfV(\sft)$ represents the fuel volume.

Since the shell is of infinitesimal thickness, we identify the fuel volume $\sfV(\sft)$ with the volume inside the cavity of the shell, which is formally given by $\sfV(\sft) = \int_\mathsf{D} \, \dm^3 \vec{\mathsf{x}}$.  Here $\mathsf{D} = \mathsf{D}(\sft)$ denotes the domain inside the shell cavity.  When substituting the identity $(1/3) \del\cdot(\mathsf{r} \vec{e}_\mathsf{r})  = 1/(3\mathsf{r}^2) \pd_\mathsf{r}(\mathsf{r}^3) = 1$ into the integrand and using the divergence theorem, we can convert the volume integral into a surface integral so that $\sfV(\sft) = (1/3) \int \sfR(\vec{e}_\mathsf{r} \cdot \vec{n})\, \dm^2 \sfS$.  After substituting \Eqs{eq:basic:dS} and \eq{eq:basic:n}, we obtain a closed expression for the fuel volume:
\begin{equation}
	\sfV(\sft) = \frac{2 \pi}{3} 
					\int_0^\pi \sfR^3 \sin(\Theta) \, 
						\frac{\pd \sfT}{\pd \vartheta}\, \dm \vartheta.
	\label{eq:basic:Volaux}
\end{equation}
In the case of a symmetric spherical shell with $\sfR=\sfR(\sft)$ and $\sfT=\vartheta$, we obtain $\sfV = 4 \pi\sfR^3 /3$, as expected.

%%%%%%%%%%%%%%%%%%%%%%%%%%%%%%%%%%%%%%%%%%%%%%%%%
%%%%%%%%%%%%%%%%%%%%%%%%%%%%%%%%%%%%%%%%%%%%%%%%%
%%%%%%%%%%%%%%%%%%%%%%%%%%%%%%%%%%%%%%%%%%%%%%%%%
\section{Variational principle}
\label{sec:VP}

%%%%%%%%%%%%%%%%%%%%%%%%%%%%%%%%%%%%%%%%%%%%%%%%%
\subsection{Dimensionless variables}
\label{sec:nondim}

For the remainder of this work, it will be convenient to introduce the following dimensionless quantities:\cite{foot:serif}
\begin{equation}
\begin{aligned}
	t 		&\doteq \frac{\sft}{R_0/|\mathsf{U}_0|},\quad  &
	R 		&\doteq \frac{\sfR}{\sfR_0} 	,\quad	\\
	\UR 	&\doteq \frac{\sfUR}{|\mathsf{U}_0|}, \quad&
	\UT 	&\doteq \frac{\sfUT}{|\mathsf{U}_0|} , \\
	\sigma 	&\doteq \frac{\sfsig}{\M/(4 \pi \sfR_0^2)} , \quad &
	\T		& \doteq \sfT.
\end{aligned}
	\label{eq:nondim:var}
\end{equation}
where $\sfR_0$ and $\mathsf{U}_0$ denote the \textit{unperturbed} radius and radial velocity of the imploding shell prior to deceleration, respectively.  By \textit{unperturbed}, we mean the radius and radial velocity corresponding to the case of a perfectly symmetric 1D imploding sphere.  In \Eq{eq:nondim:var}, the normalized areal density $\sigma$ is defined as the ratio of $\sfsig$ and the total shell mass $\M$ defined in \Eq{eq:basic:M} divided by the surface area of an unperturbed sphere of radius $R_0$.  To keep the notation consistent, we define $\T=\sfT$ since $\sfT$ is already dimensionless.

%%%%%%%%%%%%%%%%%%%%%%%%%%%%%%%%%%%%%%%%%%%%%%%%%
\subsection{Phase-space Lagrangian}
\label{sec:Lagr}

One way to construct well-controlled asymptotic approximations for dynamical systems is to directly approximate a variational principle $\delta \Lambda=0$ from which the exact equations can be derived.\cite{Ruiz:2015bz,Ruiz:2015dv,Ruiz:2017ij,Ruiz:2017et,burby2020,ruiz2020,ruiz2022}  In analogy to the classical phase-space Lagrangian $\mathsf{L}= \vec{\mathsf{P}}\cdot \dot{\vec{\sfX}}-\mathsf{H}(\sft,\vec{\sfX},\vec{\mathsf{P}})$ for point particles, we may write the action $\Lambda$ for our imploding-sphere system as follows:
\begin{equation}
	\Lambda \doteq \int_0^t L[ R,\T,\UR,\UT] \, \dm t.
	\label{eq:VP:action}
\end{equation}
Here $L=L[ R,\T,\UR,\UT]$ is the Lagrangian of the system, which is a functional of $R(t,\vartheta)$, $\Theta(t,\vartheta)$, $\UR(t,\vartheta)$, and $\UT(t,\vartheta)$.  The Lagrangian is written in terms of two components:
\begin{equation}
	L = L_{\rm sym} - H,
	\label{eq:VP:L}
\end{equation}
where $L_{\rm sym} = L_{\rm sym}[ R,\T,\UR,\UT]$ is the symplectic part given by
\begin{equation}
	L_{\rm sym} \doteq 	\int_0^\pi \left( \sigma_0 \frac{\pd^2 S_0}{\pd \vartheta \pd \phi} \right) 
						\left( \UR \frac{\pd}{\pd t} R + \UT R \frac{\pd}{\pd t} \T \right) \, \dm \vartheta
	\label{eq:VP:Lsym}
\end{equation}
and $H = H[ R,\T,\UR,\UT]$ is the Hamiltonian part.  In \Eq{eq:VP:Lsym}, $\sigma_0(\vartheta) \doteq \sigma(0,\vartheta)$ is the normalized initial areal density and $\pd^2 S_0/(\pd \vartheta \pd \phi)$ is the normalized initial surface-area coefficient.  The Hamiltonian $H$ is divided into a kinetic component and a potential component:
\begin{equation}
	H = H_{\rm kin} + H_p,
	\label{eq:VP:H}
\end{equation}
where
\begin{gather}
	H_{\rm kin} \doteq \int_0^\pi  \frac{1}{2} \left( \sigma_0 \frac{\pd^2 S_0}{\pd \vartheta \pd \phi} \right)
									\left( \UR^2 + \UT^2 \right) \, \dm \vartheta, 
			\label{eq:VP:Hkin}	\\
	H_p \doteq  \frac{\Phi}{\gamma-1} \left( \frac{V^\gamma(0)}{V^{\gamma-1}(t) } \right) ,
			\label{eq:VP:Hp} \\
	V(t) \doteq \frac{1}{3} \int_0^\pi R^3 \sin(\T) \frac{\pd \T}{\pd \vartheta}  \, \dm \vartheta.
		\label{eq:VP:V}
\end{gather}
Here we introduced the dimensionless parameter
\begin{equation}
	\Phi \doteq \frac{2 \pi \sfp_0 \sfR_0^3}{\M \mathsf{U}_0^2/2},
	\label{eq:VP:Phi}
\end{equation}
which we call the ICF parameter.  $\Phi$ is proportional to the ratio of the characteristic internal energy $\mathsf{E}_{\rm int} \doteq \sfp_0 (4 \pi \sfR_0^3/3)/(\gamma-1)$ of the fuel volume and the characteristic kinetic energy $\mathsf{E}_{\rm kin} \doteq \M \mathsf{U}_0^2/2$ of the shell prior to deceleration.\cite{betti2002,christopherson2018a,hurricane2020,christopherson2020}  This parameter is analogous to the preheat parameter introduced in \Refs{Schmit:2020jd,Ruiz2023} and is usually small $\Phi \ll 1$ since $\mathsf{E}_{\rm int} \ll \mathsf{E}_{\rm kin}$.  As we shall discuss in \Sec{sec:back}, the ICF parameter $\Phi$ plays a key role in determining the convergence ratio and the fall-line characteristics of the imploding shell.

A few remarks are worth mentioning.  First, in the variational principle \eq{eq:VP:action}, the functions $\sigma_0 \pd^2 S_0/(\pd \vartheta \pd \phi)$ and $V_0$ should be considered as independent functions of $\vartheta$, \textit{not} functionals of the initial conditions $R_0$ and $\T_0$.  In other words, $\sigma_0 \pd^2 S_0/(\pd \vartheta \pd \phi)$ and $V_0$ should be considered as independent of $R$ and $\T$.  Second, we could have written the Lagrangian \eq{eq:VP:L} in terms of the radial and polar-angular canonical momenta
\begin{equation}
	\PR \doteq \sigma_0(\vartheta) \frac{\pd S_0}{\pd \vartheta \pd \phi} \UR, \qquad
	\PT \doteq \sigma_0(\vartheta) \frac{\pd S_0}{\pd \vartheta \pd \phi} R \UT.
	\label{eq:VP:P}
\end{equation}
This latter choice of dynamical variables leads to a slightly more complicated quasilinear analysis in \Sec{sec:quasi}.  For this reason, we use instead the variables $\UR$ and $\UT$ in \Eq{eq:VP:action}.

%%%%%%%%%%%%%%%%%%%%%%%%%%%%%%%%%%%%%%%%%%%%%%%%%
\subsection{Euler--Lagrange equations}

Following \App{app:var}, when invoking the least action principle $\delta \Lambda =0$ and taking variations of the action \eq{eq:VP:action}, we obtain the following Euler--Lagrange equations (ELEs):
\begin{align}
	\delta \UR \colon & \quad 
		\frac{\pd}{\pd t} R = \UR, 
		\label{eq:VP:ELE_R} \\
	\delta \UT \colon & \quad 
		\frac{\pd}{\pd t} \T = \frac{\UT}{R}, 
		\label{eq:VP:ELE_T} \\
	\delta R \colon & \quad 
		\left( \sigma_0 \frac{\pd^2 S_0}{\pd \vartheta \pd \phi} \right)  
		\frac{\pd}{\pd t} \UR = \Phi p  R^2 \sin(\T) \frac{\pd \T}{\pd \vartheta} \notag \\
		& \qquad \qquad \qquad \qquad ~~~~~~~
								+ \left( \sigma_0 \frac{\pd^2 S_0}{\pd \vartheta \pd \phi} \right) \frac{\UT^2}{R}, 
		\label{eq:VP:ELE_PR} \\
	\delta \T \colon & \quad 
		\left( \sigma_0 \frac{\pd^2 S_0}{\pd \vartheta \pd \phi} \right) \frac{\pd}{\pd t} \left( R \UT \right) 
				= - \Phi p  R^2 \sin(\T) \frac{\pd R}{\pd \vartheta},
		\label{eq:VP:ELE_PT}
\end{align}
where 
\begin{equation}
	p(t) \doteq \frac{V^\gamma(0)}{V^\gamma(t)}
\end{equation}
is the normalized pressure.  These equations are complemented by the initial conditions:
\begin{gather}
	R(0,\vartheta) 		= R_0(\vartheta) ,  \\
	\T(0,\vartheta) 		= \T_0(\vartheta) , \\
	\UR(0,\vartheta) 	= U_{R,0}(\vartheta) ,	\\
	\UT(0,\vartheta) 	= U_{\T,0}(\vartheta) , \\
	\sigma(0,\vartheta)	=	\sigma_0 (\vartheta) , \\
	\frac{\pd^2 S_0}{\pd \vartheta \pd \phi}(\vartheta)
			= 	\bigg[ \left( R_0 \frac{\pd \T_0}{\pd \vartheta} \right)^2 
							+  \left( \frac{\pd R_0}{\pd \vartheta} \right)^2 
					\bigg]^{1/2} R_0 \sin(\T_0).
	\label{eq:VP:init}
\end{gather}
To completely specify the system of equations, we introduce the boundary conditions\cite{foot:boundary}
\begin{gather}
	\frac{\pd R}{\pd \vartheta}(t,0)		= 0 ,  \qquad
	\frac{\pd \T}{\pd \vartheta}(t,0)		= 1.
	\label{eq:VP:boundary}
\end{gather}

Equations \eq{eq:VP:ELE_R}--\eq{eq:VP:boundary} constitute a nonlinear model describing an imploding spherical thin shell undergoing RT instabilities.  Apart from the thin-shell approximation and the assumption of adiabatic compression of the fuel, these equations are exact.  After substituting \Eqs{eq:nondim:var}, one can readily verify that the ELEs \eq{eq:VP:ELE_R}--\eq{eq:VP:ELE_PT} are equivalent to \Eqs{eq:basic:Raux}--\eq{eq:basic:Vaux}.

%%%%%%%%%%%%%%%%%%%%%%%%%%%%%%%%%%%%%%%%%%%%%%%%%
\subsection{Conservation laws}
\label{sec:conservation}

Equations \eq{eq:VP:ELE_R}--\eq{eq:VP:boundary} admit two important conservation laws.  First, since there are no explicit time-dependent terms appearing in the Hamiltonian, the dynamical system is autonomous so the total energy is conserved:
\begin{equation}
	H = H(t=0) = \const,
	\label{eq:VP:Hcons}
\end{equation}
where the Hamiltonian is given by \Eqs{eq:VP:H}--\eq{eq:VP:V}.  Second, since the fluid--shell system is not subject to external potentials or forces, the total momentum of the system is also conserved.  By the assumed azimuthal symmetry, the momentum components $P_x$ and $P_y$ oriented along the $x$ and $y$ axes in a Cartesian coordinate system are null.  However, the momentum component 
\begin{equation}
	P_z  \doteq 
		\int_0^\pi 	\left( \sigma \frac{\pd^2 S}{\pd \vartheta \pd \phi} \right) 
				\left[ \UR \cos(\T) - \UT \sin(\T) \right] \, 
				\dm \vartheta
	\label{eq:VP:Pzcons}
\end{equation}
is not necessarily zero and is a conserved quantity: $P_z(t)=P_z(0)$.  Both conservation laws \eq{eq:VP:Hcons} and \eq{eq:VP:Pzcons} can be verified by direct substitution of \Eqs{eq:VP:ELE_R}--\eq{eq:VP:ELE_PT}.

%%%%%%%%%%%%%%%%%%%%%%%%%%%%%%%%%%%%%%%%%%%%
%%%%%%%%%%%%%%%%%%%%%%%%%%%%%%%%%%%%%%%%%%%%
%%%%%%%%%%%%%%%%%%%%%%%%%%%%%%%%%%%%%%%%%%%%
\section{Unperturbed shell implosion}
\label{sec:back}

In this section, we discuss the case of a symmetric imploding sphere in the absence of RTI perturbations.  For an unperturbed shell, the dynamical variables are parameterized as $R = \barR(t)$, $\T=\vartheta$, $\UR = \barUR(t)$, $\UT=0$, and $\sigma_0=1$.  When substituting into \Eq{eq:VP:L}, we obtain the 1D (unperturbed) Lagrangian $\barL= \barL[\barR,\barUR]$ of the system such that
\begin{gather}
	\barL \doteq  \barL_{\rm sym} - \barH,
	\label{eq:back:L}
\end{gather}
where
\begin{equation}
	\barL_{\rm sym} \doteq 2 \barUR \frac{\dm }{\dm t} \barR,
	\qquad	\qquad
	\barH \doteq \barUR^2 + \frac{\Phi}{\bar{R}^2},
	\label{eq:back:H}
\end{equation}
are the unperturbed symplectic part and the unperturbed Hamiltonian, respectively.  Here we used $\gamma=5/3$ for the polytropic index of the compressed fluid.  From now on, we shall restrict our discussion to this particular case.  The corresponding ELEs are
\begin{align}
	\delta \barUR \colon & \qquad 
		\frac{\dm }{\dm t} \barR = \barUR, 
		\label{eq:back:ELE_R} \\
	\delta \barR \colon & \qquad 
		\frac{\dm }{\dm t} \barUR = \frac{\Phi}{\barR^3} ,  
		\label{eq:back:ELE_PR}
\end{align}
which are complemented by the normalized initial conditions $\barR(0) = 1$ and $\barUR(0) = -1$.  

\begin{figure}
	\includegraphics[scale=0.44]{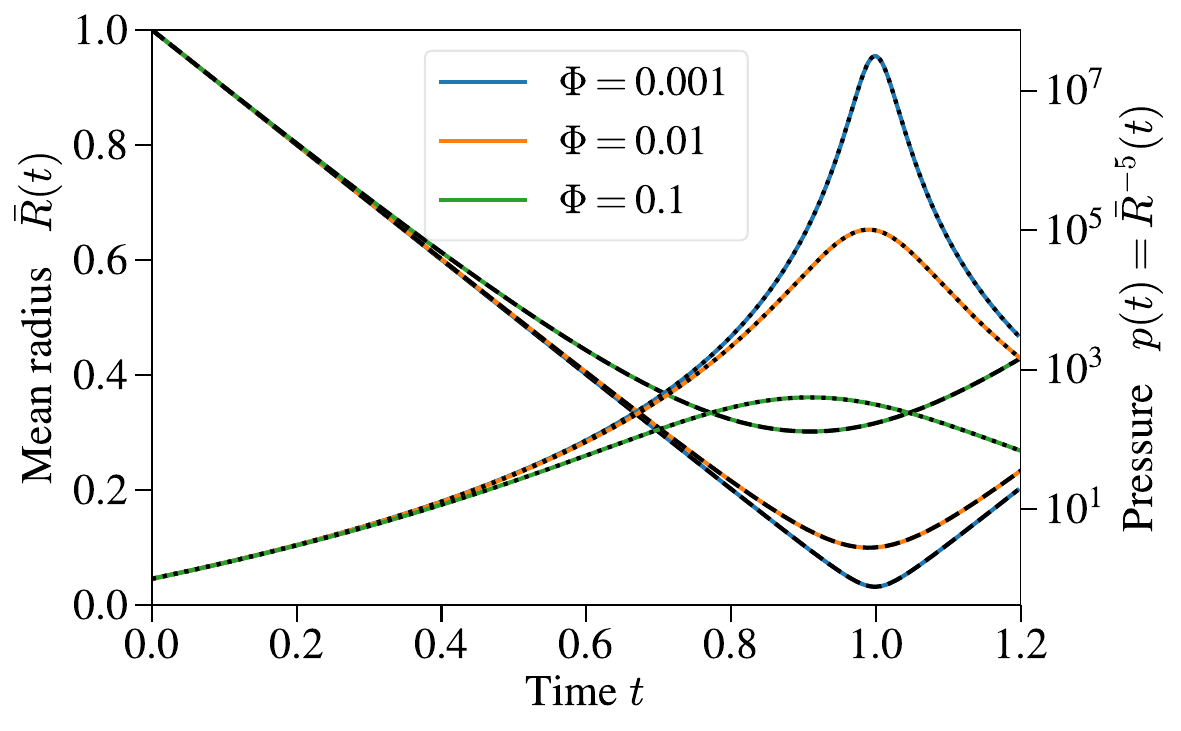}
	\caption{Comparison of the analytical solution \eq{eq:back:Rsol} to numerical integration of \Eqs{eq:VP:ELE_R}--\eq{eq:VP:boundary}.  Solid lines represent implosion trajectories in radius and pressure obtained via numerical integration.  The line colors correspond to the $\Phi$ values considered in this example.  Dashed black lines represent the analytical solution \eq{eq:back:Rsol}.  Dotted lines correspond to the analytically calculated pressure.  As shown, smaller values in the $\Phi$ parameter lead to later stagnation times, higher convergence ratios and peak pressures at stagnation, and shorter confinement times in the 1D case.}
	\label{fig:1D}
\end{figure}

Based on energy conservation \eq{eq:VP:Hcons}, we have $\barH = \barH(t=0) = 1 + \Phi$, where we substituted the initial conditions into \Eq{eq:back:H}.  Hence, the normalized radius $\bar{R}_{\rm 1D,stag}$ at stagnation, which occurs when the kinetic energy is zero, is given by
\begin{equation}
	\bar{R}_{\rm 1D,stag}
		\doteq 	\frac{\sfR_{\rm 1D,stag}}{\sfR_0} 
		= 		f^{1/2}(\Phi)  \simeq \Phi^{1/2},
	\label{eq:back:Rstag}
\end{equation}
where we introduced the function
\begin{equation}
	f(\Phi) \doteq \frac{\Phi}{1+\Phi}.
	\label{eq:back:fPhi}
\end{equation}
In \Eq{eq:back:Rstag}, the approximate expression on the right-hand-side is taken in the $\Phi\ll1$ limit.  We also added the ``1D" subscript to $\bar{R}_{\rm 1D,stag}$ to emphasize that this is the stagnation  radius for the 1D implosion case.  (As discussed in \Sec{sec:quasi:Rbar}, $\barR_{\rm stag}$ will have a small positive shift in the presence of RTI.)  Similarly, the normalized mean areal density is given by $\barsig = \bar{R}^{-2}$, so the normalized areal density at stagnation is
\begin{equation}
	\barsig_{\rm 1D,stag} 
		\doteq \frac{\sfsig_{\rm 1D,stag}}{\M/(4 \pi \sfR_0^2)}
		=  f^{-1}(\Phi) = \Phi^{-1}.
	\label{eq:back:sigma}
\end{equation}
Concerning the normalized pressure $p_{\rm stag}$ at stagnation, we obtain
\begin{equation}
	p_{\rm 1D,stag} \doteq \frac{\sfp_{\rm 1D,stag}}{\sfp_0} 
					= \frac{1}{\bar{R}_{\rm 1D,stag}^5}
					= f^{-5/2}(\Phi) \simeq \Phi^{-5/2}.
	\label{eq:back:pressure}
\end{equation}
Thus, for smaller $\Phi$ values, the shell converges to smaller radii, and the areal density and pressure increase with decreasing $\Phi$ parameter.

To calculate the characteristic confinement time during the stagnation event, let us first discuss the 1D solution for the implosion trajectory.  When inserting \Eq{eq:back:ELE_R} into the energy conservation equation $[\barH = 1 + \Phi]$ and integrating, we find that the 1D implosion trajectory is given by
\begin{equation}
	\bar{R}(t) 
		= R_{\rm 1D,stag} 
		\left [ 1 + \frac{(t-t_{\rm 1D,stag})^2}{\Phi^{-1}f^2(\Phi)} \right]^{1/2},
	\label{eq:back:Rsol}
\end{equation}
where $t_{\rm 1D,stag}\doteq 1/(1+\Phi)$ is the 1D stagnation time.  The solution \eq{eq:back:Rsol} was previously reported in \Refs{betti2001,hurricane2016a,christopherson2020}.  Figure~\ref{fig:1D} compares the analytical solution in \Eq{eq:back:Rsol} to the numerical solution of \Eqs{eq:VP:ELE_R}--\eq{eq:VP:boundary} initialized with the 1D initial conditions stated at the beginning of this section.  This comparison serves as a verification test of the numerical integrator used in this study, whose details can be found in \App{app:numerics}.

We define the confinement time $\Delta t$ as the full-width, half-maximum (FWHM) time interval of the pressure curve.  When substituting \Eq{eq:back:Rsol} into $\smash{p_{\rm 1D,stag}/2 =  [\bar{R}(\Delta t/2)]^{-5}}$ and solving for $\Delta t$, we find
\begin{equation}
	\Delta t_{\rm 1D} 
		\doteq \frac{\Delta \sft_{\rm 1D}}{\sfR_0/|\mathsf{U}_0|}
		= [4(2^{2/5}-1)]^{1/2} \, \Phi^{-1/2} f(\Phi) \simeq 1.13 \Phi^{1/2},
	\label{eq:back:tau}
\end{equation}
where $\Delta \sft_{\rm 1D}$ is the confinement time written in dimensional units.  As expected, implosions with smaller characteristic $\Phi$ values have shorter confinement times.  It is worth noting that the definition above for the confinement time (based on the FWHM of the pressure curve) is well approximated by the expression
\begin{equation}
	\Delta t \simeq 1.96	\sqrt{\frac{V_{\rm stag}}{\ddot{V}_{\rm stag}} },
	\label{eq:back:tau2}
\end{equation}
where $V_{\rm stag}$ is the fuel volume at stagnation and $\ddot{V}_{\rm stag}$ is its second derivative.  Finally, based on \Eqs{eq:back:pressure} and \eq{eq:back:tau}, we readily obtain an expression for the pressure--confinement-time parameter:
\begin{equation}
	p_{\rm 1D,stag} \Delta t_{\rm 1D} 
		\simeq 1.13 \Phi^{-1/2} f^{-3/2}(\Phi) \simeq 1.13 \Phi^{-2},
	\label{eq:back:Ptau}
\end{equation}
which increases as $\Phi^{-2}$ for small $\Phi$ values.

The stagnation metrics given in \Eqs{eq:back:pressure}, \eq{eq:back:tau}, and \eq{eq:back:Ptau} can be written in terms of quantities that are more accessible by experimental measurements and inferences, for example, in terms of the shell radius $\sfR_{\rm 1D,stag}$ at stagnation, the areal density $\sfsig_{\rm 1D,stag}$ at stagnation, and the maximum implosion velocity $\mathsf{U}_0$.  Solving for the initial gas pressure $\sfp_0$ in \Eq{eq:back:sigma} and using $\smash{\M= \sfsig_{\rm 1D,stag} 4 \pi \sfR_{\rm 1D,stag}^2}$ gives
\begin{equation}
	\sfp_0 \simeq  \frac{\sfsig_{\rm 1D,stag} \mathsf{U}_0^2  }{\sfR_0} 
					\Phi^2 ,\qquad
	\Phi 	\simeq 	\frac{ \sfR_{\rm 1D,stag}^2}{\sfR_0^2},
	\label{eq:back:expressions}
\end{equation}
where we used the $\Phi\ll 1$ limit.  When substituting \Eqs{eq:back:expressions} into \Eqs{eq:back:pressure}, \eq{eq:back:tau}, and \eq{eq:back:Ptau}, we find
\begin{gather}
	\sfp_{\rm 1D,stag}	
		\simeq \frac{\sfsig_{\rm 1D,stag} \mathsf{U}_0^2 }{\sfR_{\rm 1D,stag}}, 
		\label{eq:back:expressions2}\\
	\Delta \sft_{\rm 1D} 
		\simeq 1.13 \frac{\sfR_{\rm 1D,stag}}{|\mathsf{U}_0|}, \\
	\sfp_{\rm 1D,stag} \Delta \sft_{\rm 1D} 
		\simeq 1.13 \,\sfsig_{\rm 1D,stag} |\mathsf{U}_0 |.
		\label{eq:back:expressions4}
\end{gather}
The expressions above are written in dimensional units and are only valid in the ICF limit $(\Phi\ll 1)$.  One example on how these equations can be interpreted is the following: \Eq{eq:back:expressions4} suggests that, to increase the pressure--confinement-time parameter $\sfp_{\rm stag} \Delta \sft$, it is necessary to assemble fuel conditions with high areal density and high implosion velocity.  Since $\smash{\sfsig_{\rm 1D,stag} \propto \M/\sfR_{\rm 1D,stag}^2}$, \Eq{eq:back:expressions4} indicates that the $\smash{\sfp_{\rm 1D,stag} \Delta \sft_{\rm 1D}}$ metric is closely related to the shell radial momentum, which is proportional to $\M \mathsf{U}_0$, prior to deceleration.

To conclude this section, it is worth mentioning some obvious limitations of the present thin-shell model from the kinematics point of view.  First, this model overestimates the compression work during the deceleration phase on the adiabatic fluid. This occurs because the model does not account for the energy needed to compress the shell itself.  Second, after the stagnation event occurs, this model can overpredict the confinement time since the model assumes that all the shell mass is participating in the inertial confinement of the expanding adiabatic fluid.  In reality, not all mass participates in the confinement due to the transitory shocks within the shell.\cite{betti2001,betti2002,christopherson2018,christopherson2018a,springer2018}  Hence, the equations above must be considered as estimates only, which could be improved with the addition of new physics.

%%%%%%%%%%%%%%%%%%%%%%%%%%%%%%%%%%%%%%%%%%%%
%%%%%%%%%%%%%%%%%%%%%%%%%%%%%%%%%%%%%%%%%%%%
%%%%%%%%%%%%%%%%%%%%%%%%%%%%%%%%%%%%%%%%%%%%
\section{Nonlinear calculations of RTI growth}
\label{sec:nonlinear}

Equations \eq{eq:VP:ELE_R}--\eq{eq:VP:boundary} cannot be analytically solved for general initial conditions.  In this section, we illustrate results of numerical simulations of \Eqs{eq:VP:ELE_R}--\eq{eq:VP:boundary}, which were solved using the numerical algorithm detailed in \App{app:numerics}.  For all cases shown, the interface was discretized in 200 points in the polar angle, and the numerical timestep was set to $\delta t=0.001$.  

\begin{figure*}
	%\includegraphics[scale=.32]{fig01_polar_Lmode0}
	%\vspace{-0.1cm}
	%\includegraphics[scale=.32]{fig02_polar_Lmode1}
	%\vspace{-0.1cm}
	%\includegraphics[scale=.32]{fig03_polar_Lmode2}
	%\vspace{-0.1cm}
	%\includegraphics[scale=.32]{fig04_polar_Lmode4}
	\includegraphics[scale=0.76]{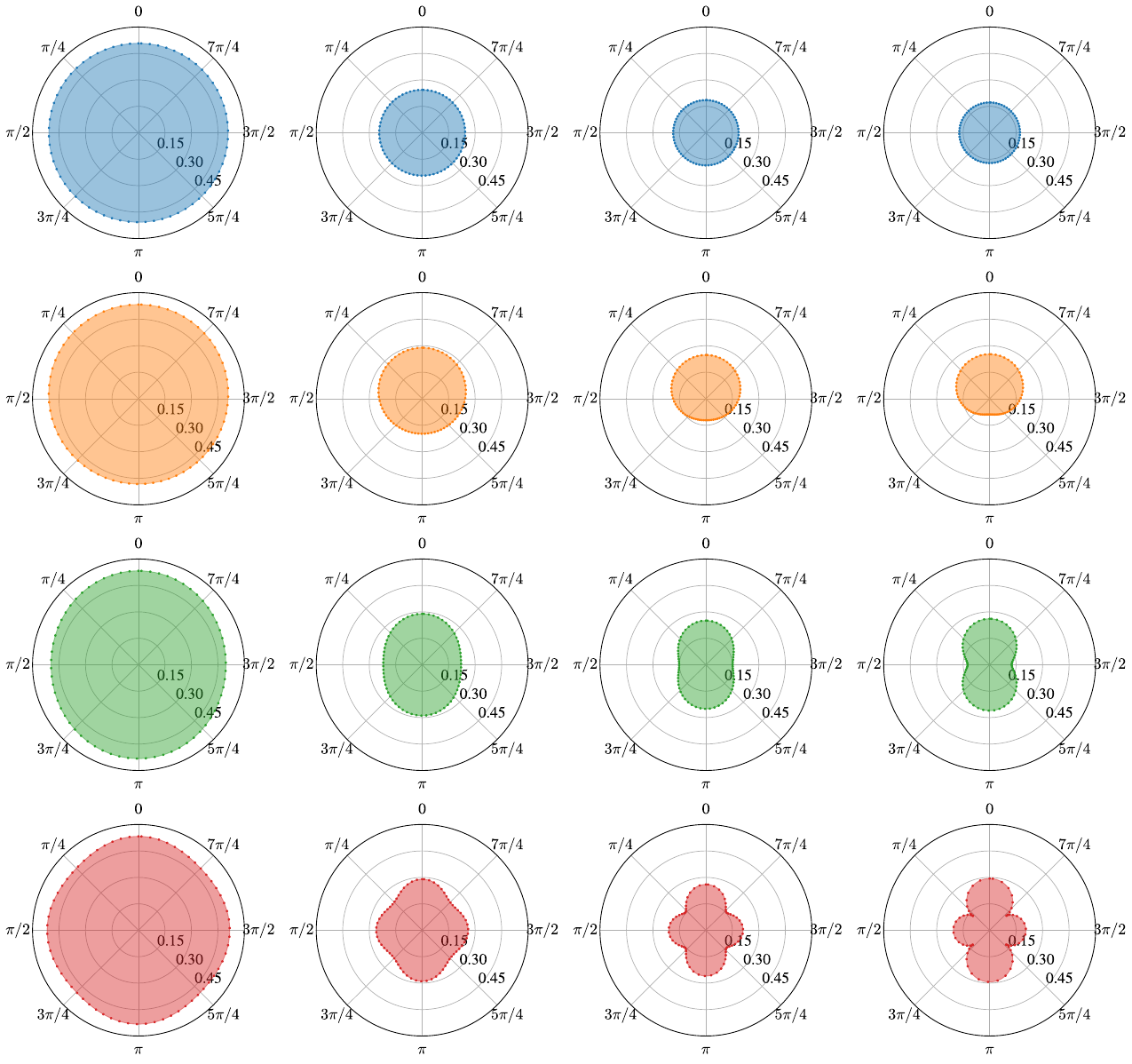}
	\caption{Temporal dynamics of the RTI for the case of a initial symmetric imploding shell with a radial velocity perturbation.  Rows from top to bottom correspond to the (i) unperturbed (symmetric) implosion, (ii) Legendre mode $\ell=1$, (iii) Legendre mode $\ell=2$, and (iv) Legendre mode $\ell=4$, respectively.  Columns correspond to times $t=\{0.5,0.8,0.9,0.95\}$.  Dots denote the position of the nodes discretizing the surface (not all are shown).  Shaded regions denote the cavity volume inside the shell.}
	\label{fig:RTI_snapshots}
	\vspace{-0.3cm}
\end{figure*}

We discuss how the degradation of the stagnation conditions depends on the Legendre mode $\ell$ of the RTI perturbations.  We consider the case of parameter $\Phi=0.03$, which leads to a moderate convergence ratio of $\mathrm{CR}_{\rm 1D,stag} \doteq \barR_{\rm 1D,stag}^{-1} \simeq 5.9$ for the 1D implosion case.   We assume that the shell is initially spherically symmetric with no areal-density perturbations so that $R_0=1$, $\Theta_0=\vartheta$, and $\sigma_0=1$.  The initial polar velocity is considered null ($U_{\Theta,0}=0$), and the initial radial implosion velocity is perturbed so that
\begin{equation}
	U_{R,0}(\vartheta) = -1 + \hatUR(0) P_\ell\boldsymbol{(}\cos(\vartheta) \boldsymbol{)},
	\label{eq:nonlinear:URpert}
\end{equation}
where $P_\ell(x)$ denotes the Legendre polynomial of degree $l$, and $\smash{\hatUR(0)}$ is the initial amplitude of the radial-velocity perturbations and is set to $\hatUR(0)=0.05$ for $\ell\neq0$.

Figure~\ref{fig:RTI_snapshots} presents a sequence of snapshots of the spherical implosions with the preimposed radial-velocity perturbations.  In the top row, the unperturbed implosion case (shown in blue) remains symmetric throughout the implosion.  For the perturbed cases, shell regions that are initially imploding slower converge less and develop RTI-bubble regions of lower areal density.  In contrast, shell regions that are initially imploding faster converge more and develop RTI-spike regions of higher areal density.  As an example, for the $\ell=1$ mode (shown in the second row in orange), the initial perturbation \eq{eq:nonlinear:URpert} causes the north pole of the sphere to implode slower than the south pole.  In consequence, the last frame of the second row in \Fig{fig:RTI_snapshots} shows a higher concentration of Lagrangian nodes in the south pole, indicating a region of higher areal density. As expected from the perturbation in \Eq{eq:nonlinear:URpert}, the $\ell=1$ case clearly shows a drift in the positive axial direction.  It is worth noting that this model produces a minimum in areal density in the direction of the residual hot-spot velocity, a correlation that is commonly observed in ICF experiments.\cite{rinderknecht2020,mannion2021b}  We shall discuss this point more quantitatively in \Sec{sec:quasi:Uz}.  Regarding the $\ell=2$ perturbation (shown in the third row in green), the imploding sphere develops bubble regions of low areal density near the poles, while spike structures of high areal density develop on the waist of the sphere.  This particular shape of the shell is reminiscent of Fig.~(9) in \Refa{kritcher2014}, where the impacts of $\ell=2$ drive asymmetries where studied in ICF implosions.  Finally, the $\ell=4$ mode (shown the fourth row in red) develops a large bubble features in the pole regions and secondary bubbles along the equator; this is caused by the shape of the $P_4(x)$ function.  In \Refa{scott2013}, the effects of $\ell=4$ mode asymmetries were discussed.  The numerical calculations of shell asymmetries shown in that work had similar shapes as this example.

\begin{figure}
	\includegraphics[scale=.42]{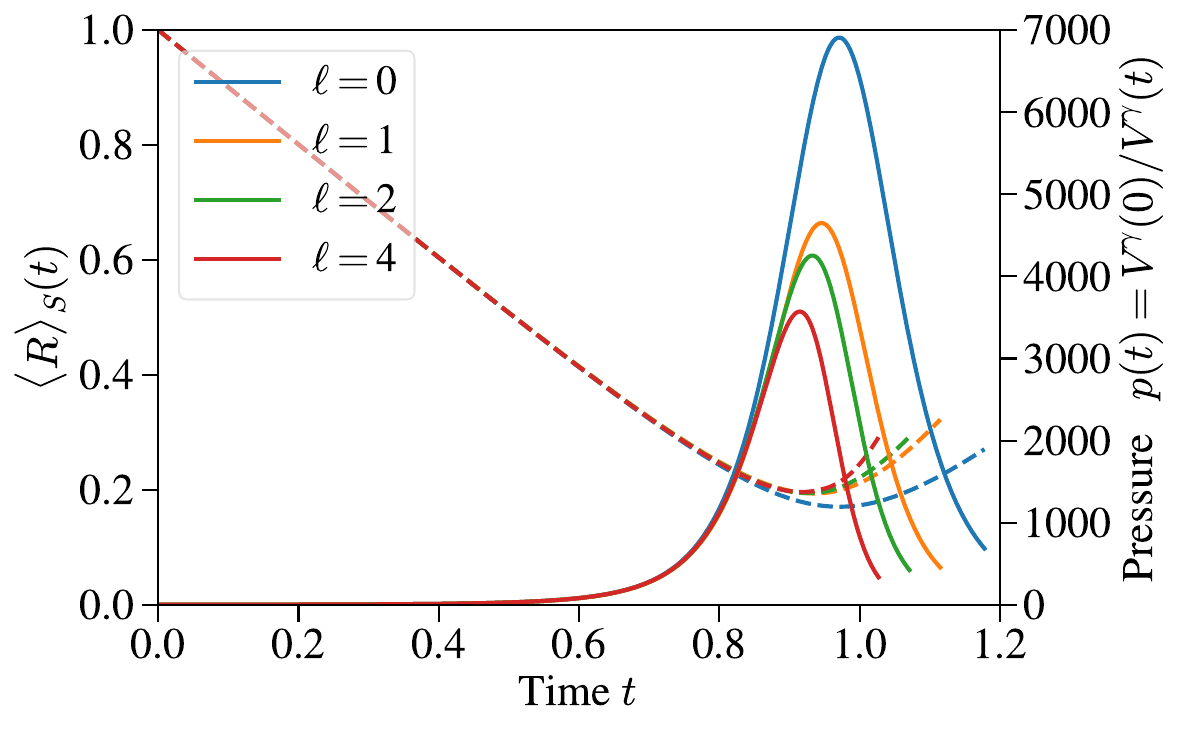}
	\caption{Implosion trajectories for the surface-weighted mean radius (dashed lines) and pressure (solid lines) versus time for different Legendre mode cases.}
	\label{fig:P_R}
	\vspace{-0.5cm}
\end{figure}

Figure~\ref{fig:P_R} compares the timetraces of the implosion trajectory of the surface-weighted mean radius $\langle R \rangle_S(t)$ and the normalized pressure for the four cases shown in \Fig{fig:RTI_snapshots}.  The surface-weighted mean radius $\langle R \rangle_S(t)$ is defined as
\begin{equation}
	\langle R \rangle_S(t) \doteq \frac{\int R \, \dm^2 S}{\int\dm^2 S},
	\label{eq:nonlinear:Ravg}
\end{equation}
where $\dm^2 S$ is given by the nondimensional version of \Eq{eq:basic:dS}.  For the cases considered here of $\Phi=0.03$ and $\smash{\hatUR(0)=0.05}$, the RTI can reduce the peak pressure by roughly $50\%$ for the $\ell=4$ mode.  The pressure timetraces also show a reduction in the FWHM time interval $\Delta t$.  This result contradicts the typical 1D paradigm of lower stagnation-pressure implosions having longer confinement times for fixed shell mass.\cite{Hurricane:2019hd}  This observation can be explained in two manners.  First, from a global perspective, the growth rate of the RTI modes increases with the shell acceleration, which peaks near stagnation.  Thus, a rapidly growing RTI mode near stagnation will increase the cavity volume and cause a reduction in pressure. Second, from a local viewpoint, the highly compressed fluid inside the shell cavity will always try to find the weak points in the confining shell near stagnation so that it can expand.  These weak points correspond to regions of low areal density (bubbles).  These regions therefore expand faster (than the unperturbed scenario) and lead to a faster pressure drop.   Both physical pictures agree with the observed earlier bounce of the surface-weighted mean radius trajectories shown in \Fig{fig:RTI_snapshots} after peak pressure has been reached.  Concerning the question on why the higher $\ell$ modes degrade more the stagnation event, larger RTI $\ell$ modes have larger linear growth rates as discussed in \Sec{sec:quasi}.  Thus, higher RTI $\ell$ modes grow faster and degrade more the stagnation conditions of the implosion.

Previous work in \Refs{kritcher2014,woo2018a,hurricane2020,hurricane2022} have noted that the degradation in stagnation metrics is closely related to the residual kinetic energy of the implosion at stagnation.  In this Hamiltonian model, the pressure degradation shown in \Fig{fig:P_R} is automatically attributed to the concept of residual kinetic energy that was left unconverted to potential energy during the implosion.  As discussed in \Sec{sec:conservation}, the Hamiltonian \eq{eq:VP:H} of the system is conserved: $\barH = \barH(t=0) \simeq 1 + \Phi$, where we substituted the initial conditions and ignored the contributions from the small RTI initial perturbations.  Hence, the pressure $p_{\rm stag}$ at stagnation will be given by
\begin{equation}
	1 + \Phi = H_{\rm kin,stag} + \Phi p_{\rm stag}^{2/5},
\end{equation}
where $H_{\rm kin,stag}$ is the residual kinetic energy at stagnation.  Solving for the pressure and substituting $p_{\rm 1D,stag}$ in \Eq{eq:back:pressure}, we find
\begin{equation}
	\frac{p_{\rm stag}}{p_{\rm 1D,stag}} = 
			\left( 1 - \frac{1}{1+\Phi} H_{\rm kin,stag} \right)^{5/2}.
	\label{eq:nonlinear:pstag}
\end{equation}
Therefore, any residual kinetic energy $H_{\rm kin,stag}$ will always reduce the stagnation pressure.  Equation~\eq{eq:nonlinear:pstag} is in agreement with the theoretical models proposed in \Refs{woo2018a,hurricane2020}, and is consistent with the findings discussed in \Refa{bose2017} in the absence of mass ablation.

\begin{figure}
	\includegraphics[scale=.42]{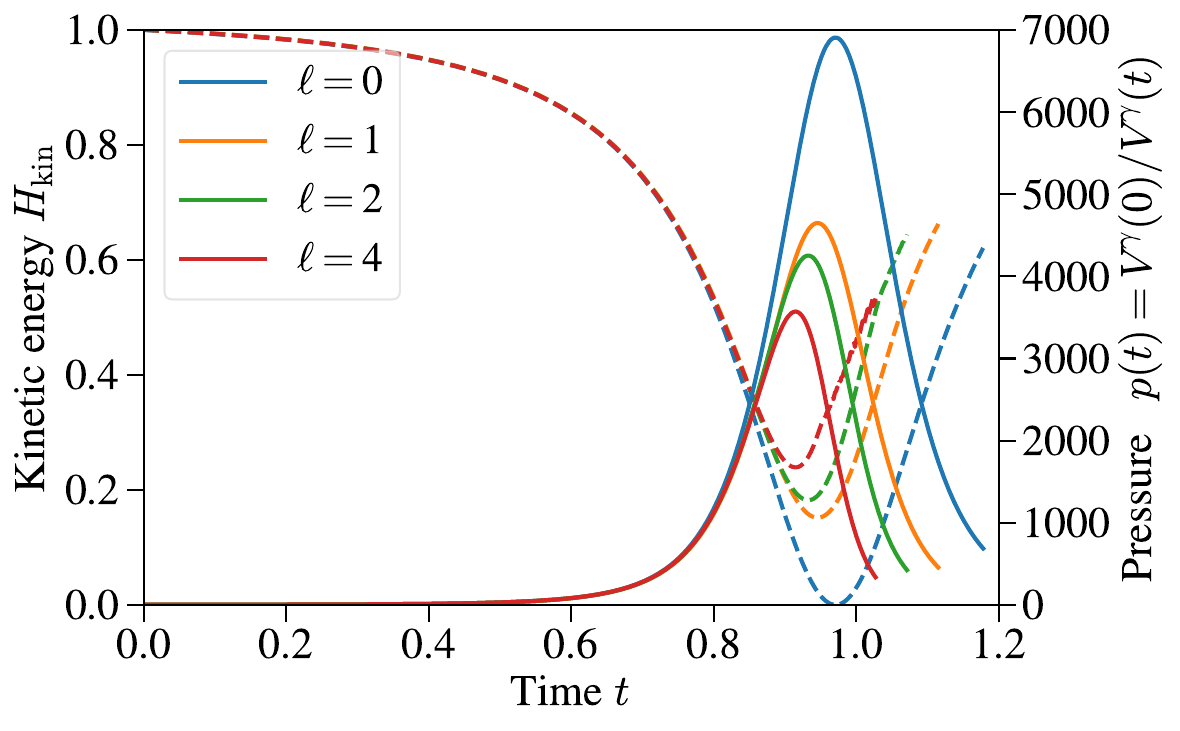}
	\caption{Shell kinetic energy (dashed lines) and pressure (solid lines) versus time for different Legendre mode cases.}
	\label{fig:P_Hkin}
	\vspace{-0.5cm}
\end{figure}

Figure~\ref{fig:P_Hkin} plots $H_{\rm kin}(t)$ for the example implosions considered, where  $H_{\rm kin}(t)$ was calculated from the simulations using \Eq{eq:VP:Hkin}.  As an example of \Eq{eq:nonlinear:pstag}, based on \Fig{fig:P_Hkin}, $H_{\rm kin,stag} \simeq 0.25$ for the $\ell=4$ mode.  Therefore, $\smash{p_{\rm stag}/p_{\rm 1D,stag} \simeq (1-0.25)^{5/2} \simeq 48\%}$, which is the observed degradation in peak pressure as compared to the unperturbed case.

It is clear then that the residual kinetic energy $H_{\rm kin,stag}$ is a fundamental quantity determining the degradation of performance metrics of interest in ICF implosions.  In the next section, we shall build a quasilinear model to understand how $H_{\rm kin,stag}$ relates to the ICF parameter $\Phi$ and to parameters characterizing RTI perturbations.

%%%%%%%%%%%%%%%%%%%%%%%%%%%%%%%%%%%%%%%%%%%%
%%%%%%%%%%%%%%%%%%%%%%%%%%%%%%%%%%%%%%%%%%%%
%%%%%%%%%%%%%%%%%%%%%%%%%%%%%%%%%%%%%%%%%%%%
\section{Quasilinear model}
\label{sec:quasi}

To gain further intuition on the dynamics described by \Eqs{eq:VP:ELE_R}--\eq{eq:VP:boundary}, we develop in this section a \textit{quasilinear} (QL) model describing the two-way nonlinear feedback between the background dynamics of the imploding sphere and the RT perturbation growth.  We shall construct this model by Taylor expanding the action \eq{eq:VP:action} up to second order in the perturbation amplitude.

%%%%%%%%%%%%%%%%%%%%%%%%%%%%%%%%%%%%%%%%%%%%
\subsection{Parameterization}
\label{sec:quasi_parameterization}

We assume that the RTI perturbations are small.  The dynamical variables are parameterized as follows:
\begin{align}
	R(t,\vartheta) 		&= \barR(t) + \ep  \tR(t,\vartheta)	,	
							\label{eq:quasi:Rtilde}\\
	\T(t,\vartheta)   	&=	\vartheta + \ep \tT(t,\vartheta) ,
							\label{eq:quasi:Ttilde}\\
	\UR(t,\vartheta) 	&= \barUR(t)  + \ep  \tUR(t,\vartheta),
							\label{eq:quasi:URtilde}\\
	\UT(t,\vartheta) 	&= \ep  \tUT(t,\vartheta), 
							\label{eq:quasi:UTtilde}\\
	\sigma(t,\vartheta)&= \barsig + \ep  \tsig(t,\vartheta).
							\label{eq:quasi:sigmatilde}
\end{align}
Here the ``barred" quantities denote the unperturbed variables (independent of $\vartheta$) while the quantities with a ``tilde" denote the RTI perturbations.  We introduce the parameter $\ep\ll1$ to denote the small amplitude of the RTI perturbations.  This parameter can be defined as the initial dimensionless amplitude of the perturbations and only serves for bookkeeping purposes for the upcoming calculations.  In addition to the small-amplitude parameterization in \eq{eq:quasi:Rtilde}--\eq{eq:quasi:sigmatilde}, we also consider eigenmodes of the RTI perturbations so that\cite{foot:RTImode}
\begin{align}
	\tR(t,\vartheta) 	&= \sumell \hatR(t)  
							P_\ell\boldsymbol{(} \cos(\vartheta) \boldsymbol{)} ,
							\label{eq:quasi:Rhat}\\
	\tT(t,\vartheta)   	&=-	 \sumell \hatT(t) 
							\frac{\dm }{\dm \vartheta}  
							P_\ell\boldsymbol{(} \cos(\vartheta) \boldsymbol{)} , 
							\label{eq:quasi:That}\\
	\tUR(t,\vartheta) 	&=  \sumell \hatUR(t)   
							P_\ell\boldsymbol{(} \cos(\vartheta) \boldsymbol{)} ,
							\label{eq:quasi:URhat}\\
	\tUT(t,\vartheta) 	&= -  \sumell \hatUT(t) 
							\frac{\dm }{\dm \vartheta} 
							 P_\ell\boldsymbol{(} \cos(\vartheta) \boldsymbol{)},
							 \label{eq:quasi:UThat}\\
	\tsig(t,\vartheta) 	&= \sumell \hatsig(t)  
							P_\ell\boldsymbol{(} \cos(\vartheta) \boldsymbol{)} .
							\label{eq:quasi:sigmahat}
\end{align}
Here $P_\ell(\mu)$ denote the Legendre polynomials satisfying the eigenvalue equation
\begin{equation}
	\frac{\dm}{\dm \mu} \left[ (1-\mu^2) \frac{\dm}{\dm \mu} P_\ell(\mu) \right] 
			= - \ell (\ell+1) P_\ell(\mu)
	\label{eq:quasi:PLegendre}
\end{equation}
and the orthogonality property
\begin{equation}
	\int_{-1}^1 P_m(\mu) P_n(\mu) \, \dm \mu = \frac{2}{2n+1} \delta_{mn},
	\label{eq:quasi:PLegendre_ortho}
\end{equation}
where $\delta_{mn}$ is the Kronecker delta.  For the sake of simplicity, we consider the case where the initial areal density $\sigma_0$ of the sphere satisfies
\begin{equation}
	\sigma_0 \frac{\pd^2 S_0}{\pd \vartheta \pd \phi} =  \sin(\vartheta),
	\label{eq:quasi:sigma0}
\end{equation}
where the expression for $\pd^2 S_0 /(\pd \vartheta \pd \phi)$ is given in \Eq{eq:VP:init}.  For general small initial perturbations, this equation can be asymptotically solved for $\sigma_0(\vartheta)$.

%%%%%%%%%%%%%%%%%%%%%%%%%%%%%%%%%%%%%%%%%%%%
\subsection{Quasilinear variational principle}
\label{sec:quasi_var}

To obtain a quasilinear model for the RTI growth, we shall follow a procedure that is commonly used to construct reduced theories of linear waves in nonhomogeneous plasmas.\cite{whitham1965,brizard2009,dodin2014a,dodin2014,Ruiz:2015bz,Ruiz:2015dv,dodin2017,Ruiz:2017ij,Ruiz:2017et}  We first substitute \Eqs{eq:quasi:Rtilde}--\eq{eq:quasi:sigma0} into the Lagrangian \eq{eq:VP:L} and Taylor expand the Lagrangian up to $\mc{O}(\ep^2)$.  Then, we use the properties \eq{eq:quasi:PLegendre} and \eq{eq:quasi:PLegendre_ortho} of Legendre functions to integrate along the reduced angle variable $\mu \doteq \cos(\vartheta)$.  This procedure leads to a theory that describes the linear dynamics of the RTI growth but also includes a nonlinear feedback to the background implosion dynamics.  Intermediate steps of these calculations are presented in \App{app:quasi}.  When following the steps above, we obtain the following quasilinear Lagrangian:
\begin{equation}
	L_{\rm QL} = \barL +  \tL,
	\label{eq:quasi:LQL}
\end{equation}
where $\barL$ is the unperturbed Lagrangian \eq{eq:back:L} for 1D implosion and
\begin{equation}
	\tL = \tL_{\rm sym} - \tH
	\label{eq:quasi:tL}
\end{equation}
is the perturbative component describing the RTI dynamics.  The symplectic part $\tL_{\rm sym}$ is given by
\begin{equation}
	\tL_{\rm sym} =  \ep^2 \sumell \frac{2}{2\ell +1}
					\left[ \hatUR \frac{\dm}{\dm t} \hatR 
							+ \ell (\ell+1)\hatUT \barR \frac{\dm}{\dm t} \hatT \right],
	\label{eq:quasi:tLsym}
\end{equation}
and the perturbative Hamiltonian is $\tH = \tH_{\rm kin} + \tH_p$, where the kinetic and potential components are given by
\begin{gather}
	\tH_{\rm kin} =  \ep^2 \sumell \frac{1}{2\ell +1}
					\left[ \hatUR^2 + \ell (\ell+1) \hatUT^2 \right], 
		\label{eq:quasi:tHkin}\\
	\tH_p = - \ep^2 \frac{\Phi}{\barR^2} 
				\sumell \frac{2}{2\ell+1} \left[ \frac{\hatR^2}{\barR^2} 
				+  \ell (\ell+1)\frac{\hatR}{\barR} \hatT  \right].
		\label{eq:quasi:tHp}
\end{gather}
Note that the dependency on $\vartheta$ is eliminated from \Eq{eq:quasi:tL}.  It is also worth noting that in $\smash{\tL}$ both background variables $\smash{(\barR)}$ and perturbative variables $\smash{(\hatR,\hatT,\hatUR,\hatUT)}$ appear together thus indicating that the theory contains a two-way dynamical coupling between the background motion and the perturbative RTI growth.  For the sake of completeness, we note that the volume \eq{eq:VP:V} of the perturbed  shell is given by
\begin{equation}
	V(t) = 
		\frac{2}{3}\barR^3 
		+ \ep^2 \sumell \frac{2 \barR^3}{2\ell +1}
			\left[ \frac{\hatR^2}{\barR^2} 
				+  \ell (\ell+1)\frac{\hatR}{\barR} \hatT  \right]
		+ \mc{O}(\ep^3).
	\label{eq:quasi:V}
\end{equation}

%%%%%%%%%%%%%%%%%%%%%%%%%%%%%%%%%%%%%%%%%%%%
\subsection{Quasilinear Euler--Lagrange equations}
\label{sec:quasi_ELE}

From the quasilinear Lagrangian \eq{eq:quasi:LQL}, we obtain the corresponding ELEs for the Legendre modes:
\begin{align}
	\delta \hatUR \colon \qquad &
				\frac{\dm }{\dm t} \hatR  = \hatUR, 
			\label{eq:quasi:hatR} \\
	\delta \hatUT \colon \qquad &
				\barR  \frac{\dm }{\dm t} \hatT  = \hatUT, 
			\label{eq:quasi:hatT}\\
	\delta \hatR \colon \qquad &
				\frac{\dm }{\dm t} \hatUR  = \frac{\Phi}{\bar{R}^{3}} 
				\bigg[ 2 \frac{\hatR}{\barR} + \ell (\ell+1) \hatT  \bigg] ,
			\label{eq:quasi:hatUR} \\
	\delta \hatT \colon \qquad &
				\frac{\dm }{\dm t}( \barR \hatUT)  
					= \frac{\Phi}{\barR^2} \frac{\hatR}{\barR} .	
			\label{eq:quasi:hatUT}			
\end{align}
In \Eqs{eq:quasi:hatUR} and \eq{eq:quasi:hatUT}, the $\Phi/\barR^3$ term corresponds to the acceleration of the 1D imploding sphere [see \Eq{eq:back:ELE_PR}], indicating acceleration-driven RTI growth.  Also, Bell--Plesset effects,\cite{bell1951,plesset1954,Velikovich:2015jl} which describe the growth of RTI due to convergence effects, are encoded in the time derivative of $\barR$ on the left-hand side of \Eq{eq:quasi:hatUT}.  

Regarding the background implosion variables, the corresponding ELEs are
\begin{align}
	\delta \barUR 	\colon \quad &
					\frac{\dm}{\dm t} \barR  =  \barUR,
			\label{eq:quasi:barR} \\
	\delta \barR 	\colon \quad &
					\frac{\dm}{\dm t} \barUR 
			 			= 	 \frac{\Phi}{\bar{R}^{3}}  
			 				+ 	\ep^2  \sum_\ell \frac{\ell (\ell+1)}{(2\ell+1) }
									\frac{\hatUT^2}{\barR}\notag \\
						&\qquad~~~~~
			 			- \ep^2 \frac{\Phi}{\bar{R}^{3}} \sum_\ell \frac{1}{2\ell+1}
				 				\bigg[ 4 \frac{\hatR^2}{\barR^2}
						+ 3 \ell (\ell+1) \frac{\hatR}{\barR} \hatT   \bigg]								 .
			\label{eq:quasi:barUR}
\end{align}
Compared to the 1D implosion model in \Eqs{eq:back:ELE_R} and \eq{eq:back:ELE_PR}, \Eq{eq:quasi:barR} for the mean radius is left unchanged.  However, the mean radial momentum equation \eq{eq:quasi:barUR} is modified by the presence of RTI perturbations.  The first term on the right-hand side of \Eq{eq:quasi:barUR} is the radial acceleration term for the unperturbed 1D case [see \Eq{eq:back:ELE_PR}].  The second term is the mean centrifugal force  associated to the angular momentum of the RTI perturbations.  This term increases as the mean radius $\barR$ decreases and provides a mechanism that resists the inward implosion.  Finally, the last term in \Eq{eq:quasi:barUR} describes the degradation of volume compression by the RTI modes.  One can deduce from \Eqs{eq:quasi:hatR}--\eq{eq:quasi:hatUT} that any initial perturbation in $\smash{\hatR}$ or $\smash{\hatUR}$ will lead to perturbations of the same sign in $\smash{\hatT}$ and $\smash{\hatUT}$.  In other words, this term reduces the outward mean force acting on the shell due to the degradation of the compression of the fluid by the RTI.

%%%%%%%%%%%%%%%%%%%%%%%%%%%%%%%%%%%%%%%%%%%%
\subsection{Eigenmode analysis}
\label{sec:quasi_eigen}

Let us now compute the RTI eigenmodes and associated instantaneous growth rate.  From \Eqs{eq:quasi:hatR}--\eq{eq:quasi:hatUT}, we write the two second-order ODEs for the RTI motion:
\begin{gather}
	\frac{\dm^2 }{\dm t^2} \hatR  = \frac{\Phi}{\bar{R}^{3}} 
				\bigg[ 2 \frac{\hatR}{\barR} + \ell (\ell+1)\hatT  \bigg] ,
	\label{eq:eigen:R2nd-a} \\
	\frac{\dm }{\dm t} \left( \barR^2 \frac{\dm }{\dm t} \hatT \right)  
					=  \frac{\Phi}{\bar{R}^{3}}  \hatR.
	\label{eq:eigen:T2nd-a}
\end{gather}
We note that \Eqs{eq:eigen:R2nd-a} and \eq{eq:eigen:T2nd-a} are identical to Eqs.~(43a) and (44a) in \Refa{Velikovich:2015jl} expressed in different variables. In \Refa{Velikovich:2015jl}, these equations were obtained by taking the thin-shell limit of a linear model describing an imploding, thick, incompressible shell undergoing RTI.  It is encouraging that the two independent approaches lead to the same system of equations \emph{for the RTI motion}.

To remove the term related to Bell--Plesset effects (\ie the term involving the time derivative of $\barR$), we introduce the change of variables $\smash{\hatT = \hatPsi/\barR}$.  This leads to 
\begin{gather}
	\frac{\dm^2 }{\dm t^2} \hatR  = \frac{\Phi}{\bar{R}^{3}} 
				\bigg[ 2 \frac{\hatR}{\barR} 
					+ \ell (\ell+1) \frac{\hatPsi}{\bar{R}}  \bigg] ,
	\label{eq:eigen:R2nd-b} \\
	\frac{\dm^2 }{\dm t^2} \hatPsi 
					=  \frac{\Phi}{\bar{R}^{3}} 
						\bigg( \frac{\hatR}{\barR} 
						+ \frac{\hatPsi}{\bar{R}} \bigg),
	\label{eq:eigen:T2nd-b}
\end{gather}
where we approximated $\dm^2 \barR / \dm t^2 \simeq \Phi/\barR^3$ in \Eq{eq:eigen:T2nd-b}.

We convert the second-order ODEs \eq{eq:eigen:R2nd-b} and \eq{eq:eigen:T2nd-b} into a set of algebraic equations by invoking the WKB approximation such that $\smash{\hatR = \hatalpha(t) \exp( \int \gamma_\ell \, \dm t )}$ and $\smash{\hatPsi = \hatbeta(t) \exp( \int \gamma_\ell \, \dm t )}$, where $\gamma_\ell=\gamma_\ell(t)$ is the growth rate of the RTI modes.  When considering only the time derivatives on the argument of the exponential, we obtain the following linear algebraic system:
\begin{equation}
	\begin{pmatrix}
   		\gamma_\ell^2 - 2\Phi/\barR^4 & - \ell(\ell+1) \Phi  /\barR^4 \\
   		-  \Phi/\barR^4  &	\gamma_\ell^2 - \Phi/\barR^4 
   \end{pmatrix}
	\begin{pmatrix}
   		\hatalpha \\
   		\hatbeta
   \end{pmatrix}
   = 
 	\begin{pmatrix}
   		0 \\
   		0 
   \end{pmatrix}  
   \label{eq:eigen:Matrix}
\end{equation}
Setting the determinant to zero leads to a RTI growth rate $\gamma_\ell$ and a RTI oscillation rate $\omega_\ell$ given by
\begin{gather}
	\gamma_\ell^2 
		=	(\ell+2) \frac{\Phi}{\barR^4},\qquad
	\omega_\ell^2 
		=	(\ell-1) \frac{\Phi}{\barR^4}.
	\label{eq:eigen:gamma}
\end{gather}
To our knowledge, this is the first time that the RTI growth rate and oscillation rate are calculated in spherical geometry within the thin-shell approximation.  $\gamma_\ell^2$ and $\omega_\ell^2$ are obtained when taking the positive and negative solutions of the bi-quadratic equation, respectively.\cite{foot:secular}

Upon drawing the analogy to the classical RTI growth rate in planar geometry $[\gamma^2_{\rm RTI} = kg]$, we identify $\Phi/\bar{R}^3$ in \Eq{eq:eigen:gamma} as the acceleration term [as expected from \Eq{eq:back:ELE_PR}] and $k_\ell \doteq (\ell+2)/\bar{R}$ as the effective RTI wavenumber, which increases as $\bar{R}$ becomes smaller.  Interestingly, the effective wavelength $\lambda_\ell\doteq 2\pi/k_\ell$ for the $\ell=1$ mode is $\lambda_1 = 2 \pi \bar{R}/3$, \ie $1/3$ the circumference of a sphere with radius $\barR$.  For high Legendre modes, the effective wavelength is $\lambda_\ell \simeq 2 \pi \bar{R}/\ell$, as expected.

To conclude the present modal analysis, the coefficients $\smash{(\hatR,\hatT,\hatUR,\hatUT)}$ of the RTI eigenmodes with exponential growth rate $\gamma_\ell$ are related by
\begin{equation}
	\hatT = \frac{1}{\ell+1} \frac{\hatR}{\barR}, \quad 
	\hatUR = \gamma_\ell \hatR, \quad
	\hatUT = \frac{\gamma_\ell}{\ell+1} \hatR,
	\label{eq:eigen:eigen}
\end{equation}
while the corresponding relations for the RTI oscillating eigenmodes are
\begin{equation}
	\hatT = -\frac{1}{\ell} \frac{\hatR}{\barR}, \quad 
	\hatUR = i\omega_\ell \hatR, \quad
	\hatUT = \frac{i\omega_\ell}{\ell} \hatR.
	\label{eq:eigen:eigen2}
\end{equation}
From these calculations, it should be emphasized that the ratio of the polar velocity and the radial velocity follows $\smash{\hatUT/\hatUR = (\ell+1)^{-1}}$ for the unstable RTI eigenmodes.  Thus, the relative magnitude of the polar flows decreases for higher $\ell$ modes.\cite{foot:residual_polar}  This conclusion is in agreement with the findings presented in \Refa{bose2017}, where it was shown that long-wavelength RTI modes introduced substantially more nonradial hydrodynamic motion as compared to the intermediate wavelength modes.

%These relations are essential for constructing solutions to initial-value problems.  For the remainder of this paper, we shall no longer discuss the RTI oscillatory modes and focus on the RTI exponentially growing modes.

To conclude this section, let us now relate the radial-perturbation coefficients $\smash{\hatR}$ to perturbations in the areal density $\sigma$.  When we substitute the initial condition \eq{eq:quasi:sigma0} into \Eq{eq:basic:sigma}, the normalized areal density satisfies
\begin{equation}
	\sigma \frac{\pd^2 S}{\pd \vartheta \pd \phi} = \sin (\vartheta).
\end{equation}
When substituting \Eqs{eq:quasi:Rtilde}--\eq{eq:quasi:sigmahat} and Taylor expanding up to $\mc{O}(\ep^2)$, we obtain
\begin{gather}
	\barR^2 \barsig = 1 + \mc{O}(\ep^2), \\
	\frac{\hatsig}{\barsig} + 2\frac{\hatR}{\barR} + \ell (\ell+1) \hatT 
		= \mc{O}(\ep).
	\label{eq:eigen:hatsigma}
\end{gather}
If we consider eigenmode perturbations satisfying \Eq{eq:eigen:eigen}, \Eq{eq:eigen:hatsigma} becomes
\begin{equation}
	\frac{\hatsig}{\barsig} + (\ell+2)\frac{\hatR}{\barR}  \simeq 0.
	\label{eq:eigen:hatsigma2}
\end{equation}
We have therefore obtained a constitutive equation relating the perturbation coefficients $\smash{\hatsig}$ of the areal density to the radial perturbation coefficients $\smash{\hatR}$.  Equation~\eq{eq:eigen:hatsigma2} shows that regions of smaller local radius tend to have larger local areal density.  Equation~\eq{eq:eigen:hatsigma2} also highlights the difficulty of achieving uniform areal density implosions in ICF: even for the idealized case of non-growing RTI perturbations $(\smash{\hatR}= \const)$, relative variations in areal density $\smash{\hatsig/\barsig}$ are expected to increase when the convergence ratio ${\rm CR} \doteq \barR^{-1}$ increases.  Thus, higher converging ICF implosions will always tend to be more susceptible to nonuniformities in areal density at fixed RTI perturbation amplitudes.

%%%%%%%%%%%%%%%%%%%%%%%%%%%%%%%%%%%%%%%%%%%%
\subsection{Residual flow velocity}
\label{sec:quasi:Uz}

Using the quasilinear model, we now obtain a relationship between the residual flow velocity and asymmetries in the areal density at stagnation.  We define the center-of-geometry (COG) position in the axial direction as
\begin{equation}
	\langle Z \rangle_S \doteq \frac{\int R \cos(\T) \, \dm^2 S}{\int \dm^2 S},
	\label{eq:quasi:Zcenter}
\end{equation}
where $\smash{\dm^2S = R\sin\T [ (\pd_\vartheta R)^2 +R^2 (\pd_\vartheta \T)^2]^{1/2} \, \dm \vartheta \dm \phi}$.  When adopting the parameterization $\smash{R=\barR+\ep\tR}$ and $\smash{\T=\barT+\ep \tT}$, we have to $\smash{\mc{O}(\ep)}$:
\begin{multline}
	R\cos\T \dm^2S 
		\simeq \bigg\lbrace \barR^3  \sin\vartheta \cos \vartheta
				+ \ep \bigg[ 3 \barR^2  \sin\vartheta \cos \vartheta \tR \\
					+ \frac{\dm}{\dm \vartheta} \left( \sin\vartheta \cos \vartheta \tT\right)
						\barR^3 \bigg] \bigg\rbrace\, \dm \vartheta \dm \phi.
\end{multline}
When integrating the above in $\vartheta$, only the second term survives.  We obtain
\begin{equation}
	\int R \cos(\T) \, \dm^2 S 
		\simeq \ep 6 \pi \barR^2 \int_0^\pi \sin\vartheta \cos \vartheta \tR \, \dm \vartheta
		= \ep 4 \pi \barR^2 \widehat{R}_1(t),
\end{equation}
where we used the orthogonality property \eq{eq:quasi:PLegendre_ortho}.  To leading order, the surface integral in the denominator of \Eq{eq:quasi:Zcenter} is $\int \dm^2 S = 4 \pi \barR^2$.  Therefore, the COG axial position is given by
\begin{equation}
	\langle Z \rangle_S 
		= \ep \widehat{R}_1(t) + \mc{O}(\ep^2).
	\label{eq:quasi:Zcenter2}
\end{equation}
To lowest order in $\ep$, the COG axial position is simply given by the $\ell=1$ mode coefficient of the radius.

We define the residual axial velocity as the time derivative of the COG axial position so that
\begin{equation}
	\langle U_z \rangle_S
		\doteq \frac{\dm \langle Z \rangle_S}{\dm t}
		\simeq \ep \frac{\dm \widehat{R}_1}{\dm t}
		\simeq \gamma_1 \widehat{R}_1
		= \ep \sqrt{\frac{3\Phi}{\bar{R}^2}} \frac{\widehat{R}_1}{\bar{R}},
	\label{eq:quasi:UZ}
\end{equation}
where we used the WKB approximation and used \Eq{eq:eigen:gamma} for the growth rate.  Then, when using \Eq{eq:eigen:hatsigma2}, we obtain
\begin{align}
	\langle U_z \rangle_{S,\mathrm{stag}} 
		&\simeq -\ep \sqrt{\frac{\Phi}{3\bar{R}^2_{\rm stag}}} 
					\bigg(
					\frac{\widehat{\sigma}_1}{\barsig} 
					\bigg)_{\rm stag}
		\notag \\
		& \simeq -\ep \sqrt{\frac{1+\Phi}{3}} 
					\bigg(
					\frac{\widehat{\sigma}_1}{\barsig_{\rm 1D}} 
					\bigg)_{\rm stag}.
	\label{eq:quasi:UZ2}
\end{align}
Since the residual flow velocity $\langle U_z \rangle_{S,\mathrm{stag}}$ is already a result of a perturbation expansion, we approximated $\barR_{\rm stag} \simeq \barR_{\rm 1D,stag} = \Phi^{1/2}/(1+\Phi)^{1/2}$.  In a similar vein, it is also possible to approximate the mean areal density with the 1D result $[\barsig \simeq \barsig_{\rm 1D}]$.

In agreement with the experiments reported in \Refs{rinderknecht2020,mannion2021b}, \Eq{eq:quasi:UZ2} shows that regions of highest areal density will be in the direction opposite to the residual-flow velocity, while the lowest areal density will be along the direction of the residual-flow velocity.  [Due to the Legendre mode decomposition, the polar-angle dependence of the areal-density perturbations follows $\widetilde{\sigma}_1(t,\vartheta)=\widehat{\sigma}_1(t) P_1\boldsymbol{(} \cos (\vartheta) \boldsymbol{)}=\widehat{\sigma}_1(t) \cos(\vartheta)$.]  It is also worth noting that \Eq{eq:quasi:UZ2} showing the correlation between the residual flow direction and asymmetries in the areal density was also derived in \Refa{hurricane2020} using an asymmetric-piston model.

%%%%%%%%%%%%%%%%%%%%%%%%%%%%%%%%%%%%%%%%%%%%
\subsection{Residual kinetic energy}

The residual kinetic energy $H_{\rm kin,stag}$ measured at stagnation is a crucial parameter determining the degradation of the stagnation conditions.  This was clearly illustrated in \Eq{eq:nonlinear:pstag} for the stagnation pressure.  The goal of this section is to rewrite the residual kinetic energy $H_{\rm kin,stag}$ in terms of quantities that can be measured in an ICF experiment at stagnation, more specifically, in terms of Legendre components of the asymmetries in the areal density.

There are two contributions to the residual kinetic energy: $H_{\rm kin,stag} = \barH_{\rm kin,stag} + \tH_{\rm kin,stag}$, which are associated with any residual velocity of the mean radial motion and of the RTI itself.  We shall first calculate the contribution from the mean motion.  We define stagnation as the moment when the potential energy is minimized; in other words, when
\begin{equation}
	\frac{\dm }{\dm t} H_p 
		\simeq \frac{\dm }{\dm t} 
				\left[ \frac{\Phi}{\bar{R}^2} 
						- \Phi \ep^2 \sumell \frac{2}{2\ell+1} 
									\left( \frac{\hatR^2}{\barR^4}  
											+ \ell (\ell+1) \frac{\hatR \hatT}{\barR^3}  
									\right)
				\right]=0,
	\label{eq:HKin:UR}
\end{equation}
where we substituted \Eqs{eq:back:H} and \eq{eq:quasi:tHp}.  We may consider \Eq{eq:HKin:UR} as a relationship for $\dm \barR / \dm t = \barUR$ at stagnation.  Without having to compute all the time derivatives, we quickly deduce 
\begin{equation}
	\bar{U}_{R, \mathrm{stag}} = \mc{O}(\ep^2)
\end{equation}
in the presence of RTI.  Hence, $\barH_{\rm kin,stag} \propto \bar{U}_{R, \mathrm{stag}}^2 = \mc{O}(\ep^4)$.  In other words, the contribution to the residual kinetic energy coming from the mean radial motion will be of $\mc{O}(\ep^4)$, which is a higher-order term that goes beyond the accuracy of the quasilinear model.  Thus, we ignore the contribution of the mean radial motion to the residual kinetic energy in this calculation.

When substituting \Eqs{eq:eigen:eigen} into \Eq{eq:quasi:tHkin}, we obtain the residual kinetic energy
\begin{align}
	H_{\rm kin,stag} 
		& \simeq \ep^2 \sumell \frac{1}{\ell+1}  \widehat{U}_{R,\ell,\mathrm{stag}}^2
			\notag \\
		& \simeq \ep^2 \sumell \frac{\ell+2}{\ell+1} \frac{\Phi}{\barR^2_{\rm stag}}  
			\bigg( \frac{\hatR}{\barR}\bigg)^2_{\rm stag}.
			\notag \\
		& \simeq \ep^2 (1+\Phi)		
			\sumell \frac{\ell+2}{\ell+1} 
			\bigg(\frac{\hatR}{\barR_{\rm 1D}}\bigg)^2_{\rm stag},
	\label{eq:quasi:Hkin2}
\end{align}
where we substituted $\barR_{\rm stag} \simeq \barR_{\rm 1D,stag}$ to lowest order.  When using \Eq{eq:eigen:hatsigma2}, we can write the residual kinetic energy in terms of perturbations in the areal density:
\begin{equation}
	H_{\rm kin,stag} 
		\simeq \ep^2 (1+\Phi)
			\sumell \frac{1}{(\ell+1)(\ell+2)} 
			\bigg( \frac{\hatsig}{\barsig_{\rm 1D}}\bigg)^2_{\rm stag}.
	\label{eq:quasi:Hkin3}
\end{equation}
It is interesting to note that \Eq{eq:quasi:Hkin3} is a generalization of the expressions for the residual kinetic energy given in \Refa{hurricane2020}, as it includes all components of the Legendre modes contributing to the asymmetry in areal density.  In \Refa{hurricane2022}, an alternative expression for the residual kinetic energy was given in terms of the harmonic-weighted averaged areal density.  Understanding the connection between the results of \Refa{hurricane2022} and this work will be left for a future publication.

%%%%%%%%%%%%%%%%%%%%%%%%%%%%%%%%%%%%%%%%%%%%
\subsection{Mean radius at stagnation}
\label{sec:quasi:Rbar}

With the residual kinetic energy, we can now calculate how the average shell radius $\barR_{\rm stag}$ changes at stagnation in the presence of RTI.  By energy conservation, we have
\begin{equation}
	1+ \Phi \simeq H_{\rm kin,stag} + H_{p,\mathrm{stag}} ,
\end{equation}
where we neglected any contributions to the total initial energy from the RTI perturbations.  (Their amplitude is initially small.)  We substitute \Eq{eq:quasi:Hkin2} for the residual kinetic energy $H_{\rm kin,stag}$ and \Eq{eq:quasi:tHp} for the potential energy.  Then, upon using the relations \eq{eq:eigen:eigen} for the RTI eigenmodes and performing simple algebraic calculations, we obtain
\begin{equation}
	\frac{\barR_{\rm stag}^2}{\barR_{\rm 1D,stag}^2}
		\simeq 1  
			+ \ep^2	\sumell
				\frac{\ell}{(2\ell+1)(\ell+1)} 
				\bigg( \frac{\hatR}{\barR_{\rm 1D}}\bigg)^2_{\rm stag}.
	\label{eq:quasi:barRstag}
\end{equation}
When using \Eq{eq:eigen:hatsigma2}, we can write the relation above in terms of perturbations in the areal density:
\begin{equation}
	\frac{\barR_{\rm stag}^2}{\barR_{\rm 1D,stag}^2}
		\simeq 1  
			+ \ep^2	\sumell
				\frac{\ell}{(2\ell+1)(\ell+1)(\ell+2)^2} 
				\bigg( \frac{\hatsig}{\barsig_{\rm 1D}}\bigg)^2_{\rm stag}.
	\label{eq:quasi:barRstag2}
\end{equation}
From \Eqs{eq:quasi:barRstag} and \eq{eq:quasi:barRstag2}, we observe that the mean radius of the shell at stagnation increases in the presence of RTI perturbations.  This effect correlates with the loss of pressure and the corresponding larger fluid volume when RTI-caused residual motion is present.  We note that this analytical result qualitatively explains the larger mean radius at stagnation that is observed in \Fig{fig:P_R} for the different $\ell$-mode perturbations.  In addition, this identified trend for $\barR_{\rm stag}$ is consistent with the result found in \Refa{hurricane2020} for the same quantity.

%%%%%%%%%%%%%%%%%%%%%%%%%%%%%%%%%%%%%%%%%%%%
\subsection{Surface-weighted averaged areal density}
\label{sec:quasi:areal}

We introduce the surface-weighted averaged areal density as follows:
\begin{equation}
	\langle \sigma \rangle_S
		\doteq \frac{\int \sigma \, \dm^2S}{\int \dm^2 S}.
	\label{eq:quasi:Avg_sigma_intro}
\end{equation}
Our goal is to calculate $\mc{O}(\ep^2)$ corrections to $\langle \sigma \rangle_S$ when RTI perturbations are present.  The numerator in \Eq{eq:quasi:Avg_sigma} is equal to $4 \pi$ due to mass conservation.  For the denominator, we follow a similar calculation as in \Sec{sec:quasi:Uz} and insert the parameterization $\smash{R=\barR+\ep\tR}$ and $\smash{\T=\barT+\ep \tT}$ into $\dm^2S$.  When using the orthogonality properties of Legendre functions and considering eigenmodes of the quasilinear system, we obtain
\begin{equation}
	\int \dm^2 S
		\simeq 4 \pi \barR^2 \bigg[ 1
		 + \ep^2 \sumell \frac{(\ell+1)(\ell+2)}{2(2\ell +1)} 
		 		\bigg( \frac{\hatR}{\barR}\bigg)^2 \bigg].
		\label{eq:quasi:Avg_sigma_dS}
\end{equation}
As expected, surface perturbations caused by RTI tend to increase the total surface area of the distorted sphere.  For a given perturbation amplitude $\smash{\hatR}$, higher mode perturbations have a larger effect on the surface area.  When substituting \Eq{eq:quasi:Avg_sigma_dS} into \Eq{eq:quasi:Avg_sigma_intro}, we obtain
\begin{equation}
	\barR^2 \langle \sigma \rangle_S
		\simeq 1 - \ep^2 \sumell \frac{(\ell+1)(\ell+2)}{2(2\ell +1)} 
		 		\bigg( \frac{\hatR}{\barR}\bigg)^2 .
\end{equation}
When using \Eq{eq:quasi:barRstag} to substitute for $\barR^2$, we find that the $\langle \sigma \rangle_S$ at stagnation is approximately given by
\begin{equation}
	\frac{\langle \sigma \rangle_{S,\mathrm{stag}}}{\barsig_{\rm 1D,stag}}
		\simeq 1 - \ep^2 \sumell \frac{(\ell+1)^2(\ell+2)+2 \ell}{2(2\ell +1)(\ell+1)} 
		 		\bigg( \frac{\hatR}{\barR_{\rm 1D}}\bigg)^2
	\label{eq:quasi:Avg_sigma2}
\end{equation}
In terms of asymmetries in the areal density, we can write the expression above as follows:
\begin{equation}
	\frac{\langle \sigma \rangle_{S,\mathrm{stag}}}{\barsig_{\rm 1D,stag}}
		\simeq 1 - 
			\ep^2 \sumell 
				\frac{(\ell+1)^2(\ell+2)+2 \ell}{2(2\ell +1)(\ell+1)(\ell+2)^2} 
		 		\bigg( \frac{\hatsig}{\barsig_{\rm 1D}}\bigg)^2_{\rm stag}.
	\label{eq:quasi:Avg_sigma}
\end{equation}
Overall, the mean areal density decreases when RTI perturbations are present.  At fixed amplitudes in the relative variation of the shell areal density $(\hatsig/\barsig=\const)$, low-mode asymmetries tend to decrease more the mean areal density as compared to high modes.

\begin{figure}
	\includegraphics[scale=0.68]{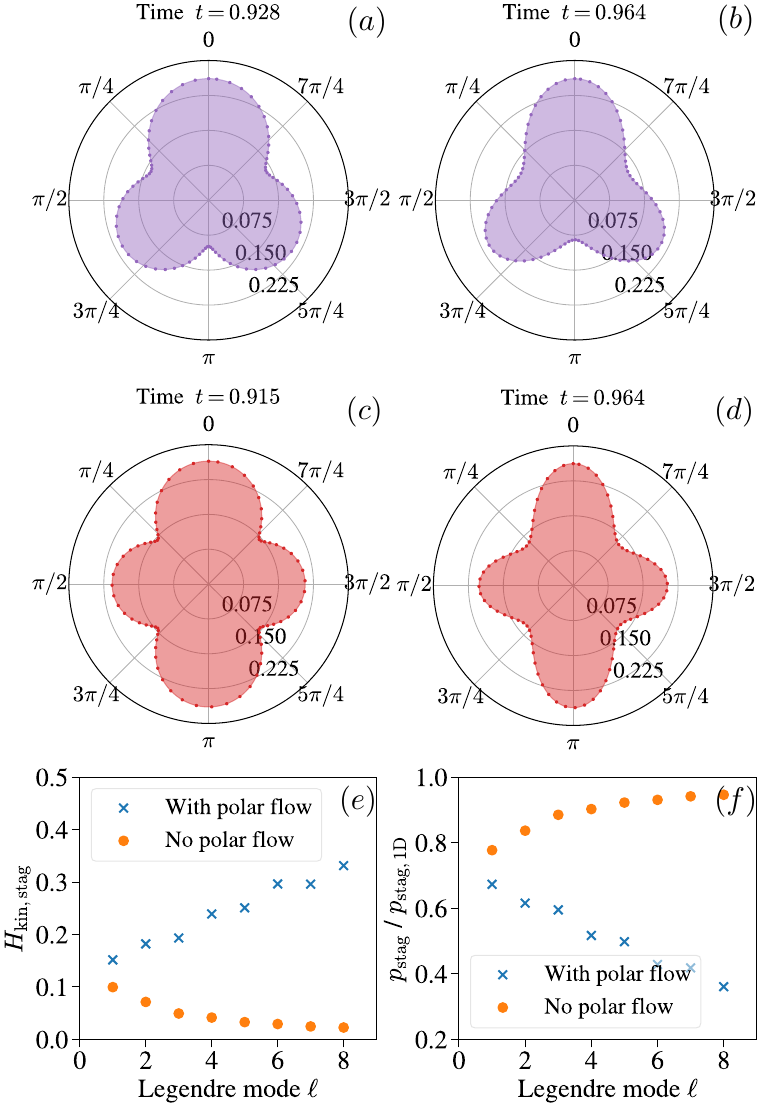}
	\caption{(a,b)~Snapshots at peak compression of seeded $\ell=3$ mode with and without polar flows, respectively. (c,d)~Snapshots at peak compression of seeded $\ell=4$ mode with and without polar flows, respectively.  (e)~Residual kinetic energy at stagnation as a function of Legendre mode $\ell$.  (f)~Degradation of the normalized pressure as a function of Legendre mode $\ell$.}
	\label{fig:Polar_noPolar}
\end{figure}

%%%%%%%%%%%%%%%%%%%%%%%%%%%%%%%%%%%%%%%%%%%%
%%%%%%%%%%%%%%%%%%%%%%%%%%%%%%%%%%%%%%%%%%%%
%%%%%%%%%%%%%%%%%%%%%%%%%%%%%%%%%%%%%%%%%%%%
\section{Discussion}
\label{sec:discussion}

In this section, we present results of nonlinear calculations of \Eqs{eq:VP:ELE_R}--\eq{eq:VP:boundary} to study the RTI dynamics in spherical implosions.  When appropriate, we shall compare our results to theoretical estimates obtained using the quasilinear model described in \Sec{sec:quasi}.

%%%%%%%%%%%%%%%%%%%%%%%%%%%%%%%%%%%%%%%%%%%%
\subsection{On the necessity of polar flows}
\label{sec:polar}

Previous models of asymmetric spherical ICF implosions have been developed where only radial motion of the shell facets is considered.\cite{springer2018}  In this section, we argue that considering both radial and polar motions is crucial not only to account for all sources of residual kinetic energy at stagnation but also to obtain the correct RTI growth during the deceleration phase.

Regarding the residual kinetic energy, suppose we have an implosion with one dominant Legendre $\ell$ mode perturbation.  Using \Eq{eq:quasi:tHkin}, we obtain the ratio of residual kinetic energy associated to radial and polar flows:
\begin{equation}
	\bigg( \frac{\tH_{\rm kin,polar}}{\tH_{\rm kin,radial}}
	\bigg)_{\rm stag}
		=	\frac{\ell (\ell+1) \widehat{U}_{\T, \mathrm{stag}}^2}{\widehat{U}_{R, \mathrm{stag}}^2}
		\simeq \frac{\ell}{\ell+1}
\end{equation}
where we used \Eq{eq:eigen:eigen} for the relationship between the velocity components.  According to this simple estimate, the radial and polar contributions to the residual kinetic energy are comparable for RTI eigenmodes.  Therefore, models that neglect flows in the polar direction can lead to an underestimation of the residual kinetic energy and the degradation of performance metrics at stagnation.

Neglecting flows in the polar direction can also lead to incorrect assessments of RTI growth.  When motion in the polar direction is neglected, the polar angle $\T$ remains constant (in time) such that $\T=\vartheta$ and $\hatT=0$.  When linearizing \Eqs{eq:VP:ELE_R} and \eq{eq:VP:ELE_PR}, the equation for the radial component $\hatR$ reads
\begin{equation*}
	\frac{\dm^2 }{\dm t^2} \hatR  = 2 \frac{\Phi}{\bar{R}^{3}}  \frac{\hatR}{\barR}.
\end{equation*}
This equation can also be retrieved by dropping the term proportional to $\hatT$ in \Eq{eq:eigen:R2nd-a}.  In this model, the instantaneous growth rate is $\gamma^2_\ell = 2 \Phi/\bar{R}^4$.  This expression 
%for the growth rate was reported in \Refa{casey2023} and 
is related to the linear growth rate in \Eq{eq:eigen:gamma} when letting $\ell=0$.  Hence, in this model with no polar flows, the RTI growth rate is independent of the $\ell$ parameter.  All Legendre modes have the same instantaneous growth rate, which is a concerning result.

To further illustrate this point, we performed a series of thin-shell calculations where the initial perturbation $\ell$ mode was changed.  We used the same simulation setup as described in \Sec{sec:nonlinear}: \ie we used $\Phi=0.03$ and considered initial radial velocity perturbations of amplitude $\smash{\hatUR(0)=0.05}$ using \Eq{eq:nonlinear:URpert}.  In \Fig{fig:Polar_noPolar}{\color{blue}(a-d)}, we compare the instability development for the $\ell=3$ and $\ell=4$ modes with and without polar flows.  When including polar flows, the bubble structures in \Fig{fig:Polar_noPolar}{\color{blue}(a,c)} are rounder and more extended, indicating higher degradation in the local areal density near the bubble regions as compared to \Fig{fig:Polar_noPolar}{\color{blue}(b,d)} with no polar flows.  

Figure \ref{fig:Polar_noPolar}{\color{blue}(e)} shows the measured residual kinetic energy $H_{\rm kin,stag}$ at peak compression calculated using \Eq{eq:VP:Hkin}.  We first note that the model with polar flows in general predicts larger residual kinetic energy.  The residual kinetic energy increases with the perturbation $\ell$ mode.  Following the results in \Sec{sec:quasi_eigen}, this is caused by an increase in the RTI growth rate for larger $\ell$ modes.  In contrast, when excluding polar flows, not only $H_{\rm kin,stag}$ is generally smaller, but it decreases with $\ell$!  Since the RTI growth remains the same for all $\ell$ modes (as argued above for this model), the decrease in $H_{\rm kin,stag}$ is likely due to the $2\ell+1$ dependency in the denominator of \Eq{eq:quasi:tHkin}.

Figure \ref{fig:Polar_noPolar}{\color{blue}(f)} shows that including polar flows leads to higher pressure degradation for large $\ell$ modes while the opposite trend occurs in the model without polar flows.  These trends in \Fig{fig:Polar_noPolar}{\color{blue}(f)} are in agreement with the relationship between residual kinetic energy and pressure degradation in \Eq{eq:nonlinear:pstag}.  These theoretical and numerical arguments illustrate the importance of including both radial and polar flows in thin-shell type models when evaluating degradation effects at stagnation due to RTI.

%%%%%%%%%%%%%%%%%%%%%%%%%%%%%%%%%%%%%%%%%%%%
\subsection{Relationship between the $\boldsymbol{\ell=1}$ mode and the residual velocity}
\label{sec:L1}

Let us now discuss how the degradation effects due to RTI depend on the ICF parameter $\Phi$ and the initial amplitude of the RTI modes.  We shall focus on the Legendre $\ell=1$ mode.  We consider the $\Phi=0.01$ and $\Phi=0.05$ cases.  According to \Eq{eq:back:Rstag}, the $\Phi=0.01$ case leads to a 1D convergence ratio of approximately $\mathrm{CR}_{\rm 1D,stag} \doteq \barR_{\rm 1D,stag}^{-1} \simeq 10$, \ie approximately double that of the $\Phi=0.05$ case.  We consider similar initial conditions for the perturbations as in \Eq{eq:nonlinear:URpert} and focus on the $\ell=1$ mode.  As in \Refa{bose2017}, we vary the initial amplitude $\smash{\hatUR(0)}$ between $0\%$ and $5\%$ to examine the degradation effects induced by RTI.

\begin{figure}
	\includegraphics[scale=0.52]{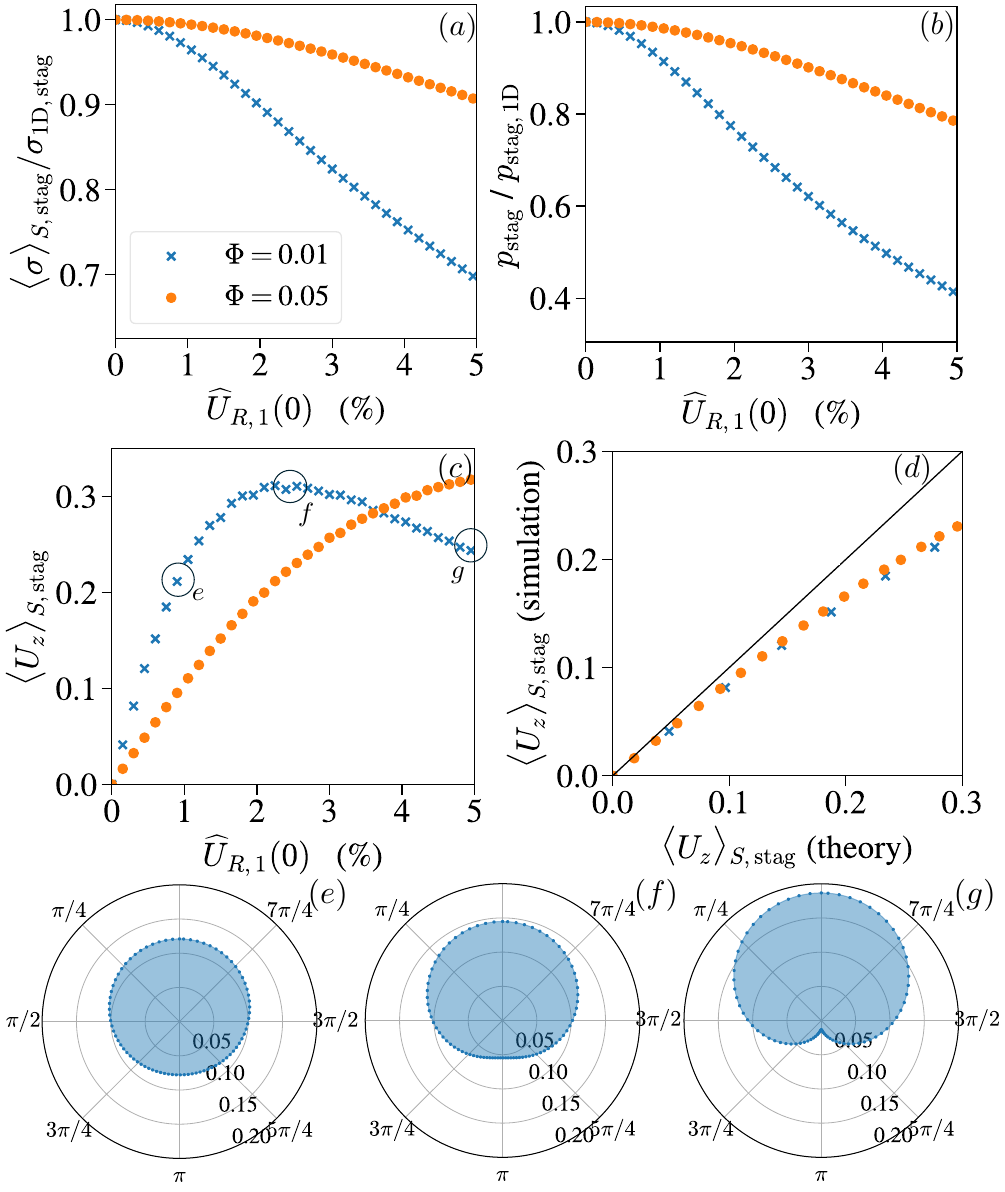}
	\caption{(a)~Degradation in the surface-weighted mean areal density as a function of the initial velocity-amplitude perturbation. (b)~Degradation of the peak pressure as a function of the initial velocity-amplitude perturbation.  For higher converging implosions with smaller $\Phi$ values, the degradation mechanisms become stronger at fixed initial amplitudes.  (c)~Axial velocity (area weighted) of the spherical shell at stagnation versus initial perturbation amplitude. (d)~Comparison of the residual flow velocity as measured from the simulations versus the theory. (e--f)~Comparison of shell asymmetries at stagnation for the cases highlighted in (c).}
	\label{fig:L1}
\end{figure}

In \Fig{fig:L1}{\color{blue}(a)}, we illustrate the degradation of the mean areal density $\smash{\langle \sigma \rangle_{S,\mathrm{stag}}}$ defined in \Eq{eq:quasi:Avg_sigma_intro}.  As the initial perturbation increases the mean areal density degrades monotonically.  A similar trend is observed for the peak pressure in \Fig{fig:L1}{\color{blue}(b)}.  In particular, the $p_{\rm stag}$ metric can decrease by $60\%$ for the $\Phi=0.01$ case at only $5\%$ in the initial mode amplitudes.  This indicates that, when pursuing implosions with higher convergence ratios, the requirements for symmetry (both in the x-ray drive and the initial shell geometry) increase.

Figure \ref{fig:L1}{\color{blue}(c)} shows the correlation between the measured residual axial velocity $\langle U_z \rangle_{S,\mathrm{stag}}$ measured at stagnation and the initial perturbation amplitude.  We define the residual axial velocity as 
\begin{equation}
	\langle U_z \rangle_S
		= \frac{\dm}{\dm t} \frac{\int R \cos(\T) \, \dm^2 S}{\int \dm^2 S}.
\end{equation}
As observed in \Fig{fig:L1}{\color{blue}(c)}, $\langle U_z \rangle_{S,\mathrm{stag}}$ is not a monotonic function of the initial $\ell=1$ perturbation amplitude.  Instead, it maximizes near a $2\%$ initial amplitude for the $\Phi=0.01$ case.  As shown by \Fig{fig:L1}{\color{blue}(e-f)}, this behavior is linked to the region in the space of initial conditions when the south polar region transitions to form a spike structure near stagnation.  According to these calculations, the residual axial velocity is maximized when the south-polar region ``flattens" at stagnation.  Once the initial perturbations are large enough to develop a spike structure near stagnation, the residual axial velocity appears to saturate since Lagrangian nodes coalescing on axis have a zero surface area.

Figure \ref{fig:L1}{\color{blue}(d)} compares the residual flow measured at stagnation against the theoretical expression \eq{eq:quasi:UZ2} relating the residual flow $\langle U_z \rangle_{S,\mathrm{stag}}$ to the $\ell=1$ component of the areal density $\sigma$.  In particular, $\widehat{\sigma}_{\rm 1,stag}$ was measured by decomposing $\sigma(t_{\rm stag},\vartheta)$ into its individual Legendre components.  We then substituted $\widehat{\sigma}_1$ into
\begin{equation}
	\langle U_z \rangle_{S,\mathrm{stag}} 
		\simeq -\ep \sqrt{\frac{\Phi}{3\barR_{\rm stag}^2}} \frac{\widehat{\sigma}_1}{\barsig_{\rm 1D}} 
		\simeq -\ep \frac{\Phi}{\sqrt{3(1+\Phi)}}\,\widehat{\sigma}_1, 
\end{equation}
where both the mean radius $\barR_{\rm stag}$ and mean areal density $\barsig_{\rm stag}$ at stagnation are evaluated using the lowest-order 1D expressions \eq{eq:back:Rstag} and \eq{eq:back:sigma}.  We note that $\langle U_z \rangle_{S,\mathrm{stag}}$ scales linearly with the $\ell=1$ mode perturbations of the areal density, which is in agreement with the observations made in ICF capsule implosions reported in \Refs{rinderknecht2020,schlossberg2021,mannion2021b}.

As shown by \Fig{fig:L1}{\color{blue}(d)}, there is reasonable agreement between the simulations and the theory at small amplitudes.  However, deviations occur once the perturbations have relatively large amplitudes, which is caused by the breakdown of quasilinear theory.  Although quasilinear theory has its limitations, this comparison illustrates how trends from the calculations can be interpreted using reduced theoretical models such as the quasilinear model presented in \Sec{sec:quasi}.

%%%%%%%%%%%%%%%%%%%%%%%%%%%%%%%%%%%%%%%%%%%%
\subsection{Dependency of stagnation degradation on the ICF parameter $\boldsymbol{\Phi}$}
\label{sec:Phi}

\begin{figure}
	\includegraphics[scale=0.38]{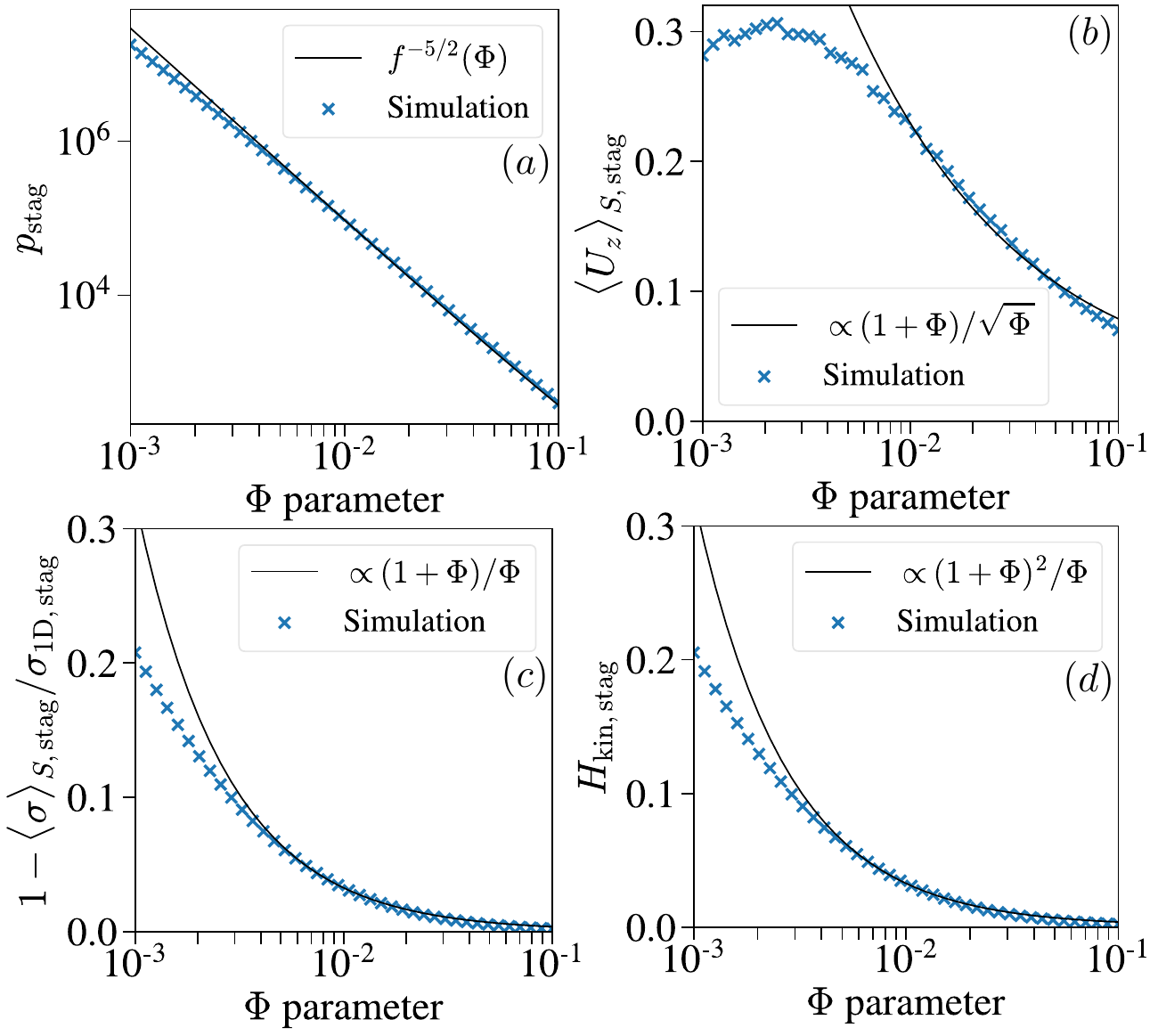}
	\caption{(a)~Pressure at stagnation, (b)~residual flow velocity, (c)~mean areal density, and (d)~residual kinetic energy as functions of the ICF parameter $\Phi$.}
	\label{fig:Phi_deg}
\end{figure}

\begin{figure}
	\includegraphics[scale=0.34]{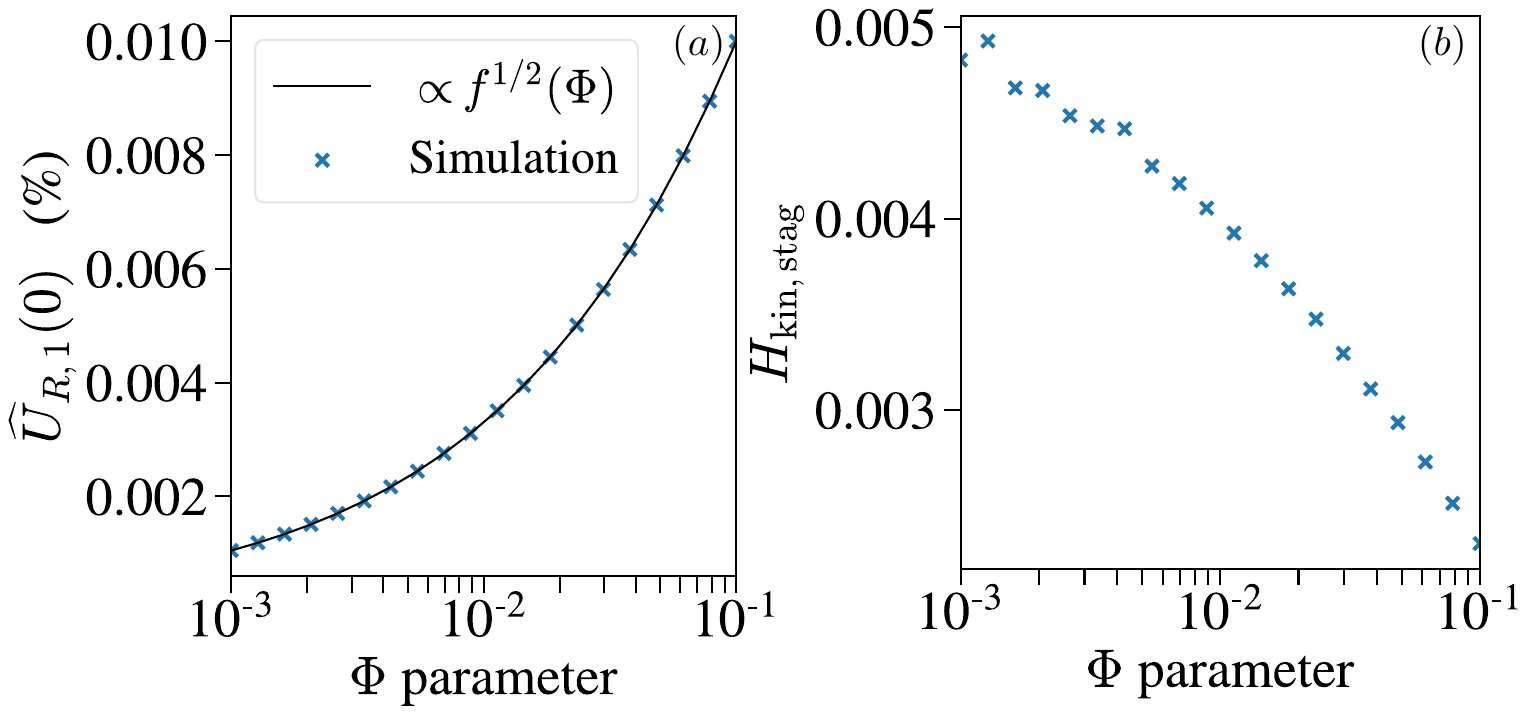}
	\caption{(a)~To maintain the relative levels of performance degradation, the initial amplitudes of the RTI perturbation must scale with the 1D stagnation radius. (b)~Measured residual kinetic energy as function of the ICF parameter $\Phi$.}
	\label{fig:Phi_same_deg}
\end{figure}

High-convergence ICF implosions are often considered more unstable.  In this section, we discuss how the degradation mechanisms due to RTI depend on the convergence ratio of an ICF implosion.  In our simplistic model, the 1D convergence ratio depends as $\mathrm{CR}_{\rm stag} \simeq \Phi^{-1/2}$ in the $\Phi \ll 1$ limit [see \Eq{eq:back:Rstag}].  This parameter will be our main variable in the study presented in this section.

Regarding the RTI, we first note that the total exponential growth of the RTI modes is only a weak function of the ICF parameter $\Phi$.  To illustrate this, the growth function $\Gamma_\ell$ for a RTI mode is $\Gamma_\ell = \int \gamma_\ell \, \dm t$, where the instantaneous growth rate is given by \Eq{eq:eigen:gamma}.  Roughly speaking, the RTI growth will scale as the product of the growth rate at stagnation and the confinement time $\Delta t$:
\begin{equation}
	\Gamma_\ell \sim \gamma_{\rm \ell, stag} \Delta t.
\end{equation}
The growth rate evaluated at stagnation is approximated by $\gamma_{\rm \ell,stag} \simeq (1+\Phi)\sqrt{(\ell+2)/\Phi} $, where we substituted \Eq{eq:back:Rstag} for the stagnated mean radius $[\barR_{\rm stag}\simeq \barR_{\rm 1D,stag}]$. The effective time period during which the RTI grows can be approximated by the confinement-time parameter $\Delta t$, which to lowest order is given by \Eq{eq:back:tau}:  $\Delta t \simeq \Delta t_{\rm 1D} \simeq 1.13 \, \Phi^{1/2}/(1+\Phi)$.  Therefore, the growth function approximately scales as\cite{book1996}
\begin{equation}
	\Gamma \sim \sqrt{(\ell+2)},
\end{equation}
which is independent of $\Phi$ to lowest order.  This simple estimate shows that the RTI growth is not necessarily greater for higher converging implosions since the increase in the growth rate $\gamma_\ell$ (due to increased acceleration) is compensated by the shorter confinement time.  Of course, higher-order effects can change these conclusions.

For high-convergence ICF implosions, increased degradation of the stagnation metrics are likely due to spherical convergence effects, not increased RTI growth.  To illustrate this, let us consider the following example.  The degradation in the mean areal density $\langle \sigma \rangle_S$ given by \Eq{eq:quasi:Avg_sigma2} scales approximately as $1- \langle \sigma \rangle_{S,\mathrm{stag}}/\barsig_{\rm 1D,stag} \propto \smash{(\widehat{R}_{\rm \ell, stag}/\barR_{\rm 1D,stag})^2}$.  Based on the discussion in the previous paragraph, we do not expect that the RTI amplitude $\smash{\widehat{R}_{\rm \ell, stag}}$ will change significantly when varying the ICF parameter at fixed initial perturbation amplitude. Therefore, the degradation in the mean areal density will primarily depend on the ICF parameter as follows: $1-\langle \sigma \rangle_{S,\mathrm{stag}}/\barsig_{\rm 1D,stag} \propto (1+\Phi)/\Phi$, where we used \Eq{eq:back:Rstag} for the mean stagnation radius.  We can repeat this logic to obtain the expected dependencies on the ICF parameter $\Phi$ for the residual velocity and the residual kinetic energy.  We summarize our results below:
\begin{gather}
	\langle U_Z \rangle_{S,\mathrm{stag}} \propto \frac{1+\Phi}{\sqrt{\Phi}},
		\label{eq:discussion:Phi:Uz}\\
	1- \frac{\langle\sigma\rangle_{S,\rm{stag}}}{\barsig_{\rm 1D,stag}} \propto \frac{1+\Phi}{\Phi},
		\label{eq:discussion:Phi:Sigma}\\
	H_{\rm kin,stag} \propto \frac{(1+\Phi)^2}{\Phi}.
		\label{eq:discussion:Phi:Hkin}
\end{gather}

In \Fig{fig:Phi_deg}, we show results of an ensemble of thin-shell calculations and compare them to the theoretical scaling laws above.  As initial conditions for the RTI perturbations, we considered a $\ell=1$ mode perturbation with initial amplitude of $\hatUR(0)=0.01$ [see \Eq{eq:nonlinear:URpert}].  The $\Phi$ parameter was varied between $[0.001,0.3]$.  In \Fig{fig:Phi_deg}{\color{blue}(a)}, we show the peak pressure as measured from the calculations and compare it to the 1D scaling curve in \Eq{eq:back:pressure}. In this model, the peak pressure monotonically increases with decreasing $\Phi$ parameter.  The calculation results follow closely the theoretical curve and only small deviations are observed (in the log scale) for the smallest $\Phi$ values.  Figure~\ref{fig:Phi_deg}{\color{blue}(b)} shows the measured residual flow velocity.  In this case, the scaling curve \eq{eq:discussion:Phi:Uz} only matches the simulation results for moderate values of the $\Phi$ parameter.  The rollover in $\langle U_Z \rangle_{S,\mathrm{stag}}$ for the smallest $\Phi$ values is likely due to the saturation effects discussed in \Sec{sec:L1}.  Finally, \Fig{fig:Phi_deg}{\color{blue}(c,d)} shows the degradation in areal density and the residual kinetic energy as measured from the calculations and compares them to the scaling curves \eq{eq:discussion:Phi:Sigma} and \eq{eq:discussion:Phi:Hkin}.  The theory reproduces well the simulation results.  Discrepancies occur for the smallest $\Phi$ values where the RTI perturbations are no longer small in amplitude and the quasilinear model begins to break down.  The results shown in \Fig{fig:Phi_deg} partially confirm that degradation effects of RTI on higher-convergence implosions are likely due to spherical convergence effects, not increased RTI growth.

In the quasilinear theory, the figure of merit for RTI-caused degradation of the implosion performance is the parameter $\smash{\widehat{R}_{\rm \ell, stag}/\barR_{\rm stag}}$, or equivalently, $\smash{\widehat{\sigma}_{\rm \ell, stag}/\barsig_{\rm stag}}$.  Thus, to maintain relative degradation in performance across $\Phi$ values, the initial amplitudes of the RTI perturbations must scale with $\barR_{\rm stag} \simeq \barR_{\rm 1D,stag} = f^{1/2}(\Phi)$.  To confirm this hypothesis, \Fig{fig:Phi_same_deg} shows simulation results for an ensemble of calculations where the initial perturbation in the flow velocity was varied according to this scaling prescription.  In \Fig{fig:Phi_same_deg}{\color{blue}(a)}, we plot the considered initial values for the perturbations and their scaling prescription.  In \Fig{fig:Phi_same_deg}{\color{blue}(b)}, we report the measured residual kinetic energy at stagnation.  Compared to \Fig{fig:Phi_deg}{\color{blue}(d)}, the residual kinetic energy remains almost null across two orders of magnitude in the $\Phi$ parameter.  These results emphasize that, when attempting higher converging spherical ICF implosions, the requirements in the shell in-flight asymmetries will likely need to be more stringent when attempting to maintain similar relative levels of degradation due to RTI.

%%%%%%%%%%%%%%%%%%%%%%%%%%%%%%%%%%%%%%%%%%%%
\subsection{Differences between prolate and oblate $\boldsymbol{\ell=2}$ implosions in the nonlinear regime as a function of initial velocity asymmetries}
\label{sec:L2}

\begin{figure}
	\includegraphics[scale=0.4]{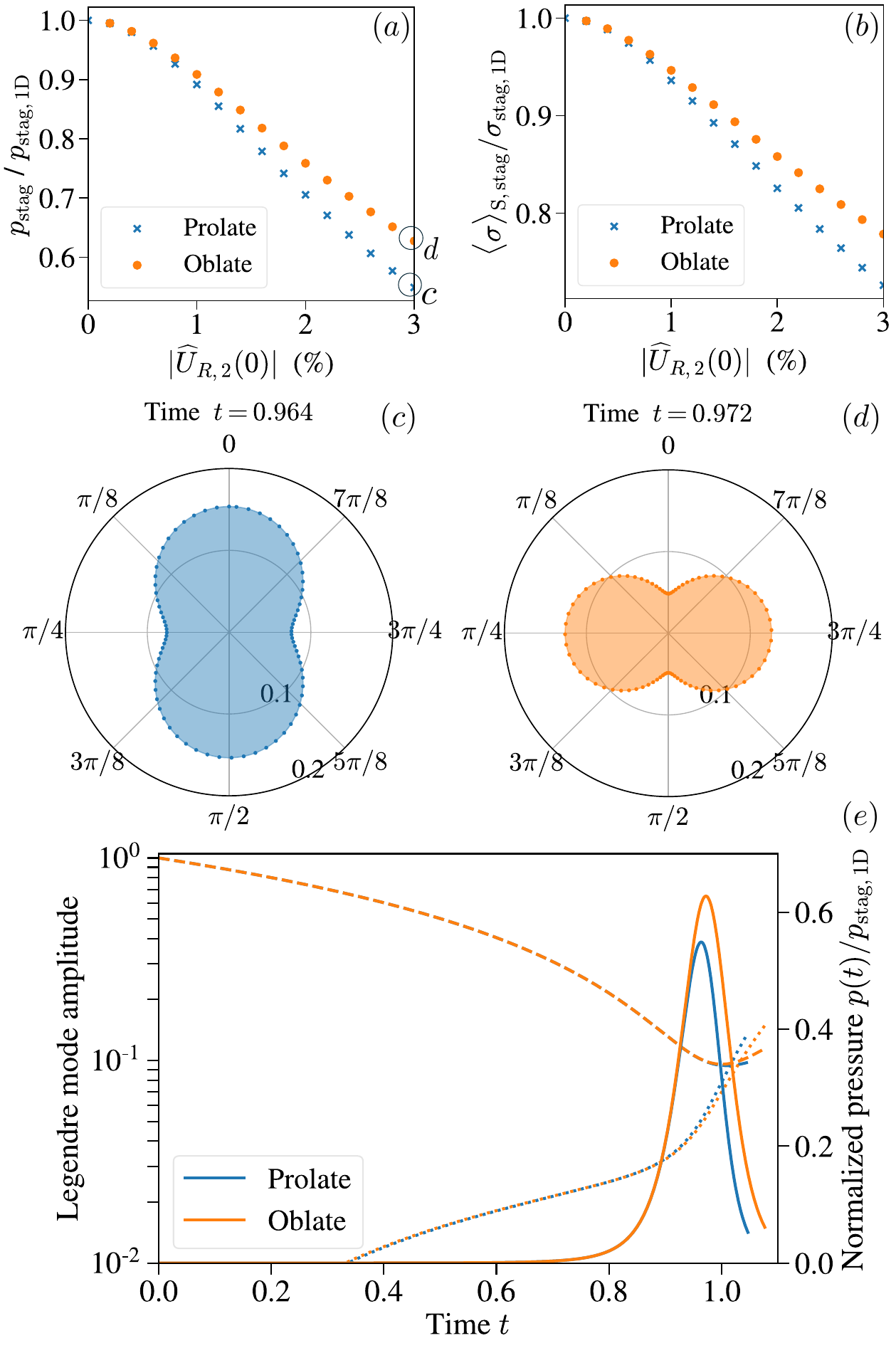}
	\caption{(a)~Degradation in the peak pressure as a function of initial velocity-amplitude perturbation. (b)~Degradation of the mean areal density as a function of initial velocity-amplitude perturbation.  (c)~Snapshot of the prolate implosion at peak compression.  (d)~Snapshot of the oblate implosion at peak compression.  (e) Temporal evolution of the mean radius (dashed line), $\ell=2$ component of the radial deformation (dotted line), and normalized pressure (solid line) corresponding to the two examples shown in (c-d).}
	\label{fig:L2_mode}
	\vspace{-0.25cm}
\end{figure}

Let us now investigate the dynamics of the $\ell=2$ mode implosions.  Figure~\ref{fig:L2_mode} presents simulations results for an ensemble of $\ell=2$ implosions.  In this study, we consider $\Phi=0.01$.  Initial perturbations in the radial velocity field follow \Eq{eq:nonlinear:URpert}, where the initial amplitude was varied between $[-0.03,0.03]$.  Negative values of the initial perturbation amplitude are considered in order to study differences in the dynamics between prolate and oblate implosions.  By prolate implosions, we mean implosions whose shell have a positive $\ell=2$ amplitude such as the one shown in \Fig{fig:Phi_deg}{\color{blue}(c)}.  Conversely, a oblate implosion has a negative $\ell=2$ amplitude as shown in \Fig{fig:L2_mode}{\color{blue}(d)}.

As shown in \Fig{fig:L2_mode}{\color{blue}(a,b)}, when increasing the absolute values of the initial velocity perturbation, the degradation of the pressure and mean areal density at stagnation increases, as expected.  For small initial amplitudes, the degradation effects of prolate and oblate implosions are equal.  This observation aligns with quasilinear theory, as degradation effects in the residual kinetic energy manifest as squared terms in $\smash{\widehat{R}_{\rm \ell, stag}/\barR_{\rm stag}}$, so implosions with positive or negative amplitude values have the same effect. The quadratic dependency of the degradation of these metrics with respect to small initial velocity perturbations is consistent with the results presented in \Refa{bose2017}.\cite{foot:bose}

%The functional dependencies of the degradation in the pressure and the areal density at stagnation shown in Fig.~10 agree with the results in Ref.~[16].

Interestingly, when the initial amplitude of the velocity perturbation is greater than 1\%, differences appear between the prolate and oblate implosions.  In fact, at fixed absolute initial amplitudes, prolate implosions show greater degradation in peak pressure and mean areal density as compared to oblate implosions.  This effect goes beyond quasilinear theory since this model is agnostic of the sign of the RTI perturbations.

It is important to note that initializing the velocity perturbation with a negative Legendre-mode amplitude does not simply correspond to ``rotating" the deformed sphere by $\pi/2$ in the polar direction. Such an interpretation could lead to the incorrect expectation that the pressure degradation is invariant with respect to the sign of the initial perturbations. There are two reasons for this.  First, we observe that $|P_2(\pm 1)|>|P_2(0)|$.  Therefore, for a fixed absolute value of the initial Legendre-mode perturbation, the bubble regions near the poles of a prolate implosion experience a greater initial local velocity perturbation compared to the bubble region near the equator of an oblate implosion. This issue can be addressed by considering velocity perturbations proportional to $P_2\boldsymbol{(}\sin(\vartheta)\boldsymbol{)}$ for oblate implosions, which effectively shifts the perturbation by $\pi/2$ in the polar angle. However, even with this adjustment, the pressure degradation does not become invariant with respect to the sign of the perturbations.  This lack of invariance arises due to the assumed axisymmetry of the shell in our 2D model. Specifically, in a 2D model, spike perturbations in a prolate implosion affect the shell along its entire waist at the equator [see \Fig{fig:L2_mode}{\color{blue}(c)}].  In contrast, spike perturbations in oblate implosions only puncture the sphere at the poles [see \Fig{fig:L2_mode}{\color{blue}(d)}].  In a 3D description of an imploding shell undergoing RTI, any rotation of the initial velocity perturbations in the polar direction should result in the same degradation of the stagnation conditions. Our current 2D model does not exhibit this rotational invariance.  Improving the present model to 3D will be the focus of future work. Nonetheless, this observation on rotational invariance could serve as a straightforward verification test for 3D radiation-hydrodynamic models\cite{clark2016,colaitis2022} that describe ICF capsule implosions.

So, what causes the observed differences shown in \Fig{fig:L2_mode} between prolate and oblate implosions?  To explain this phenomenon, \Fig{fig:L2_mode}{\color{blue}(e)} compares the temporal evolution of the mean radius and $\ell=2$ component of the radial perturbations corresponding to the two examples shown in \Fig{fig:L2_mode}{\color{blue}(c,d)}.  As shown, the trajectories of the amplitudes of the Legendre components are closely similar for both cases.  Hence, we believe that the observed differences in the pressure at stagnation are due to higher-order effects of the RTI amplitudes when calculating the volume within the shell.  To be more specific, let us consider the expression for the volume of the shell given in \Eq{eq:VP:V}.  Suppose we have a dominant $\ell=2$ eigenmode.  Then, the eigenfunctions for the radius and polar angle are given by $\smash{R = \barR + \ep\widehat{R}_2 P_2\boldsymbol{(}\cos(\vartheta)\boldsymbol{)}}$ and $\smash{\T = \vartheta - \ep(\widehat{R}_2/3) \widehat{R}_2 \pd_\vartheta P_2\boldsymbol{(}\cos(\vartheta)\boldsymbol{)}}$, where we used the parameterization presented in \Sec{sec:quasi_parameterization} and the relations in \Eq{eq:eigen:eigen} for the eigenmode coefficients.  When inserting these expressions into \Eq{eq:VP:V}, we obtain
\begin{equation}
	V = \frac{2}{3}\barR^3 
			\left( 1 
				+ \ep^2\frac{9}{5} \frac{\widehat{R}_2^2}{\barR^2} 
				+ \ep^3\frac{23}{35} \frac{\widehat{R}_2^3}{\barR^3} 
			\right) + \mc{O}(\ep^4).
\end{equation}
The first two terms are reproduced by the quasilinear result in \Eq{eq:quasi:V} when considering RTI eigenmodes.  The last term is $\mc{O}(\ep^3)$ and is absent in \Eq{eq:quasi:V}.  This term is also obviously sign dependent.  Prolate implosions with $\smash{\widehat{R}_2>0}$ tend to increase the cavity volume as compared to oblate implosions with $\smash{\widehat{R}_2<0}$.  Therefore, at fixed absolute value of the Legendre-mode amplitude, prolate implosions tend to have greater losses in pressure since these have larger cavity volumes as compared to oblate implosions.  This geometrical difference between prolate and oblate implosions explains the discrepancies in \Fig{fig:L2_mode}{\color{blue}(a)}.

%%%%%%%%%%%%%%%%%%%%%%%%%%%%%%%%%%%%%%%%%%%%
%%%%%%%%%%%%%%%%%%%%%%%%%%%%%%%%%%%%%%%%%%%%
%%%%%%%%%%%%%%%%%%%%%%%%%%%%%%%%%%%%%%%%%%%%
\section{Conclusions and future work}
\label{sec:conclusions}

In this work, we obtained a first-principle variational theory that describes an imploding spherical shell undergoing Rayleigh--Taylor instabilities (RTI).  The model is based on a thin-shell approximation and includes the dynamical coupling between the imploding spherical shell and an adiabatically compressed fluid within its interior.  Based on the derived Hamiltonian framework, degradation trends of key ICF performance metrics (e.g., stagnation pressure, residual kinetic energy, and aerial density) are identified in terms of the RTI-related parameters such as the initial amplitude and Legendre mode, as well as the implosion characteristics including the convergence ratio.  Additionally, to gain further insight into the underlying dynamics, we derive a \textit{quasilinear} model that allows us to construct explicit expressions for the residual flow velocity, mean areal density, and residual kinetic energy as functions of asymmetries in the shell areal density. Finally, we compare the results of our theoretical predictions to the numerical calculations solving the fully nonlinear equations.

This work can be expanded in several ways.  First, this model can be generalized to include 3D RTI asymmetries.  Previous work on this topic was presented in \Refs{taguchi1995,woo2018a}, but it may be warranted to revisit this problem with a special emphasis on studying the relationship between the 3D RTI eigenmodes and degradation of the stagnation conditions in ICF implosions.  Since 3D radiation-hydrodynamic calculations are computationally expensive,\cite{clark2016,colaitis2022} theoretical insights gained from such a reduced 3D, thin-shell model could be valuable.  Second, the model can be improved to describe the acceleration phase of a spherical implosion.  The current work can be expanded to incorporate mass ablation of the shell surface, as in the rocket model\cite{decoste1979,murakami1987} and in recent advancements in reduced modeling of ICF implosions.\cite{callahan2020}  This enhanced description would provide insights into the interactions between shell perturbations induced by asymmetries in the radiation drive, their subsequent acceleration (or RTI) growth, and the resulting impact on the stagnation phase of the implosion.  Third, this study specifically focused on the scenario of an adiabatically compressed fluid within the shell cavity. It would be interesting to improve the current model for the hot spot by incorporating internal heat sources (\eg alpha heating), energy losses (\eg bremsstrahlung x-ray losses), and mass ablation from the cold shell into the hot spot.  Such future body of work could be considered as an extension of the model proposed in \Refa{hurricane2021} and would provide a more detailed physical picture of the relationship between RTI and the initiation of the ignition process. A related study on this topic was presented in \Refa{woo2021a}.

%\acknowledgments

The author is indebted to O.~M.~Mannion, C.~A.~Williams, P.~F.~Schmit, D.~T.~Casey, and O.~A.~Hurricane for fruitful discussions.  Sandia National Laboratories is a multimission laboratory managed and operated by National Technology $\&$ Engineering Solutions of Sandia, LLC, a wholly owned subsidiary of Honeywell International Inc., for the U.S. DOE NNSA under contract DE-NA0003525.  

This paper describes objective technical results and analysis. Any subjective views or opinions that might be expressed in the paper do not necessarily represent the views of the U.S. DOE or the U.S. Government.

The data that support the findings of this study are available from the corresponding author upon reasonable request.

\appendix

%%%%%%%%%%%%%%%%%%%%%%%%%%%%%%%%%%%%%%%%%%%%
%%%%%%%%%%%%%%%%%%%%%%%%%%%%%%%%%%%%%%%%%%%%
%%%%%%%%%%%%%%%%%%%%%%%%%%%%%%%%%%%%%%%%%%%%
\section{Variations of the action}
\label{app:var}

In this Appendix, we illustrate how to take variations of the action with respect to dynamical variables.  Let $F\doteq F[R,\T]$ be a functional of the fields $R(t,\vartheta)$ and $\Theta(t,\vartheta)$.  A variation $\delta F/ \delta R$ of the functional $F$ with respect to the variable $R$ is defined as
\begin{equation}
	\frac{\delta F}{\delta R(t,\vartheta)} \doteq \lim_{\ep \to 0} \frac{F[R+\epsilon \delta^2(t-t',\vartheta-\vartheta'),\T]  - F[R,\T] }{\ep}, 
	\label{eq:var:var}
\end{equation}
where $\delta^2(t,\vartheta) \doteq \delta(t) \delta(\vartheta)$ and $\delta(x)$ is the Dirac delta function.  As an example, when performing the operation above on the first term in \Eq{eq:VP:Lsym}, we obtain

\begin{align}
	\frac{\delta}{\delta R(t,\vartheta)} \int 
		&\PR(t',\vartheta') \frac{\pd R}{\pd t'} \, \dm \vartheta' \, \dm t' \notag \\
		&	=  \int \PR(t',\vartheta') 
					\frac{\pd^2 \delta(t-t',\vartheta-\vartheta')}{\pd t'} \, 
						\dm \vartheta' \, \dm t' \notag \\
		&	= -  \int 	\frac{\pd \PR}{\pd t'} 
						\delta^2(t-t',\vartheta-\vartheta')\, 
						\dm \vartheta' \, \dm t' \notag \\
		&	= -  \frac{\pd}{\pd t}\PR(t,\vartheta),
\end{align}
where we integrated by parts.  Operations such as this one are repeated throughout the manuscript.  In particular, the variation of the potential component \eq{eq:VP:Hp} of the action with respect to the angle $\Theta$ gives
\begin{widetext}
\begin{align}
	\frac{\delta}{\delta \T(t,\vartheta)} \int H_p \, \dm t' 
		&	= 	-\frac{\Phi}{3}	
				\int \left( \frac{V(0)}{V(t')} \right)^\gamma
				 	R^3 \left[  \cos(\T)  
					\frac{\pd \T}{\pd \vartheta'} \delta^2(t-t',\vartheta-\vartheta')
							+ \sin(\T) \frac{\pd \delta^2(t-t',\vartheta-\vartheta')}{\pd \vartheta'} \right] \, 
					\dm t'\, \dm \vartheta' \notag \\
		&	= 	-\frac{\Phi}{3}
				\int \left( \frac{V(0)}{V(t')} \right)^\gamma
						\left[   R^3  \cos(\T) \frac{\pd \T}{\pd \vartheta'}
							- \frac{\pd}{\pd \vartheta'} \left( R^3 \sin(\T) \right)\right] \, 
						\delta^2(t-t',\vartheta-\vartheta') \, 
					\dm t'\, \dm \vartheta' \notag \\
		&	= 	 \Phi \left( \frac{V(0)}{V(t)} \right)^\gamma 
					\, R^2(t,\vartheta) 
					\sin\boldsymbol{(}\T(t,\vartheta) \boldsymbol{)}
					\frac{\pd R}{\pd \vartheta}(t,\vartheta) .
\end{align}
\end{widetext}
%
%This calculation is used for obtaining the ELE \eq{eq:VP:ELE_PT}.

%%%%%%%%%%%%%%%%%%%%%%%%%%%%%%%%%%%%%%%%%%%%
%%%%%%%%%%%%%%%%%%%%%%%%%%%%%%%%%%%%%%%%%%%%
%%%%%%%%%%%%%%%%%%%%%%%%%%%%%%%%%%%%%%%%%%%%
\section{Numerical algorithm}
\label{app:numerics}

Numerical calculations were done using a $\mc{O}( \Delta \vartheta^2 , \Delta t^2)$ symplectic integrator.  The domain $[0,\infty)\times[0,\pi]$ of the governing equations was discretized using a timestep $\Delta t>0$ and Lagrangian-label step $\Delta \vartheta>0$ so that the numerical grid is $(t^n, \vartheta_j) = (n \Delta t, j \Delta \vartheta)$ for $n\geq0$, $j \in \mathbb{Z}$.  We used a St\"{o}rmer--Verlet integrator for the time stepping.  For convenience, the equations were solved using the variables $P_R \doteq \UR$ and $P_\T\doteq R\UT$.  The algorithm is the following:
\begin{widetext}
\begin{align}
	P_{\T,j}^{n+1/2} &= P_{\T,j}^{n}  
			- \frac{\Delta t}{2} 
				\Phi p^n   
				\left[\left( \sigma_0 \frac{\pd^2 S_0}{\pd \vartheta \pd \phi} \right)_j\right]^{-1}
				 (R^n_j)^2 \sin(\T^n_j) \left(\frac{\pd R}{\pd \vartheta} \right)^n_j,
	\label{eq:app:numerics:PT}
	\\
	P_{R,j}^{n+1/2} &= P_{R,j}^{n}  
			+ \frac{\Delta t}{2} 
				\Phi p^n   
				\left[\left( \sigma_0 \frac{\pd^2 S_0}{\pd \vartheta \pd \phi} \right)_j\right]^{-1}
				 (R^n_j)^2 \sin(\T^n_j) \left(\frac{\pd \T}{\pd \vartheta} \right)^n_j
			+ \frac{\Delta t}{2}  \frac{(P_{\T,j}^{n+1/2})^2}{(R^n_j)^3},\\
	R^{n+1}_j 	&= R^{n}_j + \Delta t P_{R,j}^{n+1/2}, 
	\\
	\T^{n+1}_j 	&= \T^{n}_j 
				+ \frac{\Delta t}{2} 
					\left[ \frac{P_{\T,j}^{n+1/2}	}{\left(R^n_j \right)^2} 
							+ \frac{P_{\T,j}^{n+1/2}}{\left(R^{n+1}_j\right)^2}
					\right], \\
	P_{\T,j}^{n+1} &= P_{\T,j}^{n+1/2}  
			- \frac{\Delta t}{2} 
				\Phi p^{n+1}   
				\left[\left( \sigma_0 \frac{\pd^2 S_0}{\pd \vartheta \pd \phi} \right)_j\right]^{-1}
				 (R^{n+1}_j)^2 \sin(\T^{n+1}_j) 
				 \left(\frac{\pd R}{\pd \vartheta} \right)^{n+1}_j,
	\\
	P_{R,j}^{n+1} &= P_{R,j}^{n+1/2}  
			+ \frac{\Delta t}{2} 
				\Phi p^{n+1}
				\left[\left( \sigma_0 \frac{\pd^2 S_0}{\pd \vartheta \pd \phi} \right)_j\right]^{-1}
				 (R^{n+1}_j)^2 \sin(\T^{n+1}_j) 
				 \left(\frac{\pd \T}{\pd \vartheta} \right)^{n+1}_j
			+ \frac{\Delta t}{2}  \frac{(P_{\T,j}^{n+1/2})^2}{(R^{n+1}_j)^3}.
	\label{eq:app:numerics:PR}
\end{align}	
\end{widetext}
Derivatives on $\vartheta$ were approximated using finite central differences such that
\begin{gather}
	\left(\frac{\pd X}{\pd \vartheta} \right)^n_j
		\simeq \frac{X^n_{j+1}-X^n_{j-1}}{2\Delta \vartheta}.
\end{gather}
In a similar manner, the mass term $\sigma_0 (\pd^2S/\pd \vartheta \phi)$ was approximated using finite central differences.  The pressure term is given by $p^n = (V^0/V^n)^\gamma$, where the volume is calculated using the trapezoidal rule:
\begin{equation}
	V^n \simeq 	\frac{1}{3}
			\sum_j \frac{(R_j^n)^3\sin(\Theta^n_j)+(R_{j-1}^n)^3\sin(\Theta^n_{j-1})}{2}
					 \left( \T^n_j - \T^n_{j-1} \right).
\end{equation}

As discussed in \Refa{ott1972}, thin-shell models such as the one presented in this work are prone to develop cusp-like features once instabilities grow sufficiently.  Such cusps are undesirable as they introduce errors in the calculation of the fluid volume within the shell.  To fix this issue, the numerical algorithm used in this work verifies whether a cusp feature is forming after each timestep.  This procedure is briefly summarized.  First, the center-of-geometry axial position $\langle Z\rangle_S$ defined in \Eq{eq:quasi:Zcenter} is calculated for the shell configuration at the end of a timestep.  Second, based on the center-of-geometry position, a new polar angle $\T'$ is calculated for each fluid element composing the shell.  Then, a monotonicity test is done on the angle $\T'$.  Regions where $\T'$ ceases to be monotonically increasing indicate that the local region has begun to develop a cusp.  Once this occurs, neighboring Lagrangian nodes within this region are forced to ``inelastically collide".  Such collision conserves total momentum but not energy.  The ``collided" Lagrangian elements can only move together, and they ballistically evolve in time.  In other words, no external forces can act on them; they maintain their momentum.  Fluid elements that have not collided continue to evolve according to \Eqs{eq:app:numerics:PT}--\eq{eq:app:numerics:PR}.

%%%%%%%%%%%%%%%%%%%%%%%%%%%%%%%%%%%%%%%%%%%%
%%%%%%%%%%%%%%%%%%%%%%%%%%%%%%%%%%%%%%%%%%%%
%%%%%%%%%%%%%%%%%%%%%%%%%%%%%%%%%%%%%%%%%%%%
\section{Auxiliary results for the calculation of the quasilinear Lagrangian}
\label{app:quasi}

We substitute \Eqs{eq:quasi:Rtilde}--\eq{eq:quasi:sigmatilde} and \Eq{eq:quasi:sigma0} into the Lagrangian \eq{eq:VP:L}.  We then Taylor expand the Lagrangian up to $\smash{\mc{O}(\ep^2)}$. The resulting quasilinear Lagrangian $\smash{L_{\rm QL}=L_{\rm QL}[\barR,\barUR,\tR,\tT,\tUR,\tUT]}$ is $\smash{L_{\rm QL} \doteq \barL + \ep^2 \tL}$, where $\barL$ is the unperturbed Lagrangian \eq{eq:back:L} and $\smash{\tL \doteq \tL_{\rm sym} - \tH}$ is the perturbative component.  The symplectic part $\smash{\tL_{\rm sym}}$ is written as
\begin{equation}
	\tL_{\rm sym} \doteq \int_{-1}^1 
							\left( \tUR \frac{\pd}{\pd t} \tR(t,\mu) +
								\tUT(t,\mu)\bar{R}(t) \frac{\pd}{\pd t} \tT(t,\mu) 
							\right) \, \dm \mu ,
	\label{eq:quasi2:tLsym}
\end{equation}
where we performed a change of variables and introduced the reduced angle variable $\mu \doteq \cos(\vartheta)$.  The perturbative Hamiltonian $\tH = \tH_{\rm kin} + \tH_p$ is given by
\begin{gather}
	\tH_{\rm kin} \doteq \frac{1}{2} \int_{-1}^1 [\tUR^2(t,\mu) + \tUT^2(t,\mu) ] \, \dm \mu,
	\label{eq:quasi2:tHkin} \\
	\tH_p \doteq - \frac{\Phi}{\barR^2} \int_{-1}^1 \left[ \frac{\tR^2}{\barR^2} 
									-  \frac{\tR}{\barR} \frac{\pd}{\pd \mu} \left( \sqrt{1-\mu^2} \, \tT  
						 \right) \right]\, \dm \mu.
	\label{eq:quasi2:tHp}
\end{gather}
We can now compute the quasilinear equations of motion.  Regarding the RTI dynamics, the ELEs are
\begin{align}
	\delta \tUR \colon \quad &
				\frac{\pd}{\pd t} \tR  = \tUR, 
			\label{eq:quasi2:tR} \\
	\delta \tUT \colon \quad &
				\barR  \frac{\pd}{\pd t} \tT  = \tUT, 
			\label{eq:quasi2:tT}\\
	\delta \tR \colon \quad &
				\frac{\pd}{\pd t} \tUR  = \frac{\Phi}{\bar{R}^{3}} 
				\left[ 2 \frac{\tR}{\barR} 
						- \frac{\pd}{\pd \mu} \left( \sqrt{1-\mu^2} \, \tT  \right)
				\right] ,
			\label{eq:quasi2:tUR} \\
	\delta \tT \colon \quad &
				\frac{\pd}{\pd t}( \barR \tUT)  = \frac{\Phi}{\bar{R}^{3}} 
				\sqrt{1-\mu^2} \frac{\pd}{\pd \mu} \tR .	
			\label{eq:quasi2:tUT}			
\end{align}
If we parameterize the radial perturbative variables $\tR$ and $\tUR$ as functions of Legendre polynomials, e.g., $\tR = \sumell\hatR(t) P_\ell(\mu)$, then \Eq{eq:quasi2:tUT} suggests that $\tUT$ and $\T$ should be proportional to $\sqrt{1-\mu^2} \pd_\mu P_\ell(\mu)$ in order to eliminate the dependency on the reduced angle $\mu$.  This observation led to the parameterization adopted in \Eqs{eq:quasi:Rhat}--\eq{eq:quasi:sigmahat}. Regarding the background implosion variables, the corresponding ELEs are
\begin{align}
	\delta \barUR 	\colon \quad 
					\frac{\dm}{\dm t} \barR  = & \barUR,
			\label{eq:quasi2:barR} \\
	\delta \barR 	\colon \quad 
					\frac{\dm}{\dm t} \barUR 
			 			= 	& \frac{\Phi}{\bar{R}^{3}} 
			 					+ \ep^2 \int_{-1}^1 \frac{\tUT^2}{\barR} \, \dm \mu \notag \\
							&	- 	\ep^2 \frac{\Phi}{\bar{R}^{3}} 
				 				\int_{-1}^1 \bigg[ 4 \frac{\tR^2}{\barR^2}
									- 3\frac{\tR}{\barR} \frac{\pd}{\pd \mu} \left( \sqrt{1-\mu^2} \, \tT  
						 \right) \bigg]\, \dm \mu .
			\label{eq:quasi2:barUR}
\end{align}
When adopting the parameterization given in \Eqs{eq:quasi:Rhat}--\eq{eq:quasi:sigmahat} and using \Eqs{eq:quasi:PLegendre} and \eq{eq:quasi:PLegendre_ortho}, the integrals above can be calculated analytically to get \Eq{eq:quasi:barUR}.

%%%%%%%%%%%%%%%%%%%%%%%%%%%%%%%%%%%%%%%%%%%%
%%%%%%%%%%%%%%%%%%%%%%%%%%%%%%%%%%%%%%%%%%%%
%%%%%%%%%%%%%%%%%%%%%%%%%%%%%%%%%%%%%%%%%%%%
%\bibliographystyle{apsrev-title}
%\bibliography{RT_sphere,foot}

\begin{thebibliography}{65}
\bibitem[{foo({\natexlab{b}})}]{foot:boundary}
\bibinfo{note}{These boundary conditions are satisfied by the RTI eigenmodes
  discussed later in \Sec{sec:quasi}. These boundary conditions are also
  utilized when integrating by parts the action integral to obtain the
  quasilinear Lagrangian in \Eq{eq:quasi:LQL}.}
\bibitem[{foo({\natexlab{c}})}]{foot:RTImode}
\bibinfo{note}{This parameterization can be found by inserting
  \Eqs{eq:quasi:Rtilde}--\eq{eq:quasi:sigmatilde} into
  \Eqs{eq:VP:ELE_R}-\eq{eq:VP:ELE_PT} and linearizing the resulting equations.
  One can then deduce that the eigenmodes of the linearized system are Legendre
  functions.}
\bibitem[{foo({\natexlab{d}})}]{foot:secular}
\bibinfo{note}{The $\ell=1$ mode has a secular eigenmode with $\omega_1=0$; its
  growth is not exponential but rather algebraic. This eigenmode satisfies
  $\smash{\dm^2 \widehat{R}_1 /\dm t^2 = 0}$ and $\smash{
  \widehat{R}_1=-\widehat{\Psi}_1}$ so that $\smash{\widehat{R}_1(t) =
  -\widehat{\Psi}_1 = a_0 + a_1 t }$.}
\bibitem[{foo({\natexlab{e}})}]{foot:residual_polar}
\bibinfo{note}{Although $\smash{\hatUR}$ and $\smash{\hatUT}$ are related by
  $\smash{\hatUT/\hatUR=(\ell+1)^{-1}}$ for the unstable RTI eigenmodes, this
  relationship does not imply that polar flows can be disregarded for higher
  mode numbers when calculating the residual kinetic energy. This aspect is
  further elaborated in \Sec{sec:polar}.}
\bibitem[{foo({\natexlab{f}})}]{foot:bose}
\bibinfo{note}{Fig.~2 of Ref.~16 indicates that $\smash{\mathrm{YOC} \simeq 1 -
  a \widehat{U}_{R,2}^2(0)}$ for small initial perturbations, where
  $\mathrm{YOC}$ is the yield-over-clean ratio and $a$ is a constant.
  Additionally, the pressure degradation illustrated in Fig.~7~(b-ii) of
  Ref.~16 approximately follows the relationship $\smash{p_{\rm stag}/p_{\rm
  1D,stag}\simeq b(\mathrm{YOC}-1)+1}$, where $b$ is a constant. When
  substituting $\smash{\widehat{U}_{R,2}(0)}$ in place of $\mathrm{YOC}$, we
  obtain $\smash{p_{\rm stag}/p_{\rm 1D,stag}=1-ab \widehat{U}_{R,2}^2(0)}$,
  which reflects the same quadratic dependency for the pressure degradation
  derived from quasilinear theory and observed in
  \Fig{fig:L2_mode}{\color{blue}(a)}.}
\end{thebibliography}

\end{document}